\documentclass[10pt]{article}
\usepackage{bbm}
\usepackage{amssymb,amsmath,fullpage,times,epsf}
\usepackage{mathrsfs}
\usepackage{amssymb}
\usepackage{graphics}
\usepackage{graphicx}
\usepackage{subfigure}
\usepackage{float}
\usepackage{bm}
\usepackage{amsfonts}
\usepackage{color}

\allowdisplaybreaks
\def\beginproof{\par\noindent\textbf{Proof.}~~}
\def\endproof{\vbox{\hrule\hbox{\vrule height1.0ex\hskip1.0ex\vrule}\hrule }\par\medskip}

\newtheorem{theorem}{Theorem}[section]

\newtheorem{lemma}[theorem]{Lemma}

\newtheorem{proposition}[theorem]{Proposition}

\newtheorem{remark}[theorem]{Remark}

\def\lam{\lambda}
\def\mb{\mathbf}
\def\ra{\rightarrow}

\def\lim{\mathrm{lim}}

\def\max{\mathrm{max}}
\def\PT{$\mathcal{P}\mathcal{T}$}

\def\eref#1{(\ref{#1})}

\usepackage[square,numbers,sort&compress]{natbib}
\begin{document}
\date{}

\title{Rational solutions of the defocusing nonlocal nonlinear Schr\"{o}dinger equation: Asymptotic analysis and soliton interactions}

\author{Tao Xu$^{1,2,}$\thanks{Corresponding author, e-mail: xutao@cup.edu.cn}\,, Lingling Li$^{2}$, Min Li$^{3,}$\thanks{Corresponding author, e-mail: ml85@ncepu.edu.cn}\,, Chunxia Li$^{4}$, Xuefeng Zhang$^{3}$
 \\
{\em 1. State Key Laboratory of Heavy Oil Processing,}\\
{\em China University of Petroleum, Beijing 102249, China}\\
{\em 2. College of Science, China University of Petroleum, Beijing
102249, China} \\
{\em 3.  School of Mathematics and Physics,}\\
{\em  North China Electric Power University, Beijing 102206,
China}\\
{\em 4. School of Mathematical Sciences, Capital Normal University,}
 \\
{\em Beijing 100048, China 
} }
\maketitle
\vspace{-5mm}

\begin{abstract}

In this paper, we obtain the $N$th-order rational solutions for the defocusing nonlocal nonlinear Schr\"{o}dinger equation by the Darboux transformation and some limit technique. Then, via an improved asymptotic analysis method relying on the balance between different algebraic terms, we derive the explicit expressions of all asymptotic solitons of the rational solutions with the order $1\leq N \leq 4$. It turns out that the asymptotic solitons are localized in the straight or algebraic curves, and the exact solutions approach the curved asymptotic solitons  with a slower rate than the straight ones. Moreover, we find that all the rational solutions exhibit just five different types of soliton interactions, and  the interacting solitons are divided into two halves with each having the same amplitudes. Particularly for the curved asymptotic solitons, there may exist a slight difference for their velocities between at $t$ and $-t$ with certain parametric condition. In addition, we reveal that the soliton interactions in the rational solutions with $N\ge 2$ are stronger than those in the exponential and exponential-and-rational solutions.

\vspace{5mm}



\end{abstract}

\newpage

\section{Introduction}

In the late sixties, most known integrable evolution partial differential
equations (PDEs) are the local models~\cite{AblClark}, i.e., the solution's evolution dynamics depends on the values of the solution and its spatial and temporal derivatives only at one local point. In 2013, Ablowitz and Musslimani proposed the following nonlocal nonlinear Schr\"{o}dinger (NNLS) equation~\cite{Ablowitz1}:
\begin{align}
 i u_t(x,t) = u_{xx}(x,t) + 2 \varepsilon u(x,t)^2 u^{*}(-x,t)
\quad (\varepsilon= \pm 1)\,, \label{NNLS}
\end{align}
where $u$ is a complex-valued function of real variables $x$ and $t$, $\varepsilon=1$ and $\varepsilon=-1$ represent respectively the focusing $(+)$ and defocusing $(-)$ nonlinearity, and the asterisk denotes  complex conjugate.
Compared with the local nonlinear Schr\"{o}dinger (NLS) equation, the nonlinear term  $|u(x,t)|^2u(x,t)$ is replaced by $u(x,t)^2 u^{*}(-x,t)$, which reflects the space-reverse nonlocal coupling between $u(x,t)$ and $u^*(-x,t)$. Remarkably, Eq.~\eref{NNLS}
is integrable in the sense that it is the second member of the Ablowitz-Kaup-Newell-Segur hierarchy
with a complex reverse-space reduction~\cite{Ablowitz1}. This inspires that some integrable nonlocal
reductions (which include the space, time,
and space-time reversals and their combinations with complex conjugate) have been overlooked in the
known linear scattering problems, like the Ablowitz-Kaup-Newell-Segur~\cite{AKNS}, Kaup-Newell~\cite{KN}
and Wadati-Konno-Ichikawa~\cite{WKI} schemes.
Soon thereafter, 
a number of nonlocal integrable evolution PDEs have been identified in both one and two space dimensions as well as in discrete settings~\cite{Ablowitz2,Ablowitz3,Ablowitz4,Ablowitz6,Ablowitz7,Fokas,vector,
Sinha,DNLS,Rao,mKdV,NNWave,Lou2,YJK4,Tang,Cen,Yu,Lou3}.

As a new simple integrable model,  Eq.~\eref{NNLS} has been actively studied from different mathematical aspects, including the inverse scattering transform schemes of initial value problem with zero and nonzero boundary conditions~\cite{Ablowitz1,Ablowitz3,Ablowitz5}, Hamiltonian structures for the NNLS hierarchy~\cite{Gerdjikov}, long-time asymptotic behavior with decaying boundary conditions~\cite{Rybalkoy},
equivalent transformation between the NLS and NNLS equations~\cite{YJK1}, local well-posedness and blow-up instability of
arbitrarily small initial data~\cite{Genoud}, etc. Meanwhile, various analytical methods were used to derive wide classes of explicit solutions for both $\varepsilon= 1$ and $\varepsilon= -1$ cases of Eq.~\eref{NNLS}~\cite{Ablowitz1,Ablowitz3,Ablowitz5,Sarma,LiXu,LiXu1,HJS,Gupta1,YJK2,YJK3,YZY,Wen,Gurses,Khare,ZDJ,FBF,LML,Michor,
YChen,LiXu3,Santini,LiXu4}.
It has been shown that the focusing case possesses the bright-soliton, dark-soliton, rogue-wave and breather solutions, which may develop the blow-up behavior in finite time~\cite{Ablowitz1,Sarma,Khare,Gupta1,YJK2,YJK3,Santini,LiXu4},
whereas the defocusing case admits the nonsingular soliton solutions which
in general exhibit the pairwise soliton interactions~\cite{YZY,LiXu,LiXu1,LiXu3,Wen,Ablowitz5,FBF}.

On the other hand, several efforts have been made to establish the physical relevance of the NNLS equation. For example, Ref.~\cite{Magstr} found that Eq.~\eref{NNLS} is linked to an unconventional coupled Landau-Lifshitz system in magnetics via some gauge transformation, Ref.~\cite{AMJPA} derived Eq.~\eref{NNLS} as the quasi-monochromatic complex reduction of the cubic nonlinear Klein-Gordon, Korteweg-de Vries and water wave equations. Also, Eq.~\eref{NNLS} can be formally viewed as the linear Schr\"{o}dinger equation $ i u_t(x,t) = u_{xx}(x,t) + V(x,t)u(x,t)$, where the self-induced potential $V(x,t)=2 \varepsilon u(x,t)u^*(-x,t)$ naturally satisfies the parity-time (\PT) symmetric condition $V(x,t)=V^{*}(-x,t)$. Thus, the NNLS equation may have potential applications in the \PT~symmetric physics~\cite{PTPhys}. Within this regard, Ref.~\cite{HJS} discussed the similarity between the time-dependent complex potential generated from an exact solution of Eq.~\eref{NNLS} and the gain/loss distribution in a \PT~symmetric optical system; Moreover, Ref.~\cite{LiXu4} showed that the general stationary solutions of Eq.~\eref{NNLS} can be used to yield a wide class of complex and time-independent \PT~symmetric potentials whose associated Hamiltonians are \PT~symmetry unbroken.

It is an important concern in integrable systems to study the asymptotic behavior and interaction dynamics of multi-soliton solutions~\cite{Schiebold, Biondini,Anco}. Recently, we employed the Darboux transformation (DT) to obtain the exponential, rational, and exponential-and-rational solutions of the defocusing NNLS equation over the nonzero background, and revealed abundant soliton interaction phenomena in sharp contrast with the local NLS equation~\cite{LiXu,LiXu1,LiXu3}. In Refs.~\cite{LiXu,LiXu3}, via asymptotic analysis we showed that both the $N$-th order exponential and exponential-and-rational solutions contain  $2N$ asymptotic solitons, and they can display a variety of elastic interactions because each soliton could be of the dark or antidark type. In Ref.~\cite{LiXu1}, some qualitative analysis of the rational solution indicated that the asymptotic solitons  might be localized in the curved lines, and they may exhibit a less number of soliton interactions. However, the accurate asymptotic expressions have not yet been derived for the rational solutions with the order $N\ge 2$. Luckily, we obtained that the asymptotic solitons of the exponential-and-rational solutions are localized in some logarithmical curves by considering the dominant balance between the algebraic and exponential terms. This gives us an important hint to derive all the asymptotic solitons of the rational solution with any given order through a systematic procedure.

In comparison with the previous paper~\cite{LiXu1}, the difference of this work lies in three aspects: First, by using the $N$-fold DT and some limit technique, we give a rigorous proof on the determinant representation of the $N$th-order rational solutions. Second, we develop the asymptotic analysis method by considering the balance between different algebraic terms up to the subdominant level. As a result, we derive all the asymptotic solitons (which are localized  in the straight or algebraic curves) as well as their center trajectories for the rational solutions with the order $1\leq N \leq 4$. Following the same procedure, one can in principle make an asymptotic analysis of the rational solution with arbitrary order. Third, based on the asymptotic expressions, we show that all the rational solutions admit just five different types of soliton interactions, and  the interacting solitons are divided into two halves with each having the same amplitudes. Particularly, the curved asymptotic solitons can exhibit the  \emph{quasi-elastic} behavior~---~their velocities have a slight difference between at $t$ and $-t$ with certain parametric condition. In addition, we reveal that the soliton interactions in the rational solutions with $N\ge 2$ are stronger than those in the exponential and exponential-and-rational solutions.

The structure of this paper is organized as follows: In Section~\ref{Sec2}, we review the $N$-fold DT
as proposed in Ref.~\cite{LiXu}, and then construct the $N$-th order rational solutions of Eq.~\eref{NNLS}
with $\varepsilon= -1$ via some limit technique. In Section~\ref{Sec3}, we use an improved asymptotic analysis method to derive the explicit expressions of all asymptotic solitons of the rational solutions with the order $1\leq N \leq 4$, and also make a comparison of asymptotic solitons with the exact solutions. In Section~\ref{Sec4}, we discuss the dynamics of soliton interactions in the rational solutions, and particularly reveal some unusual soliton behaviors.  Finally in Section~\ref{Sec5}, we address the conclusions and discussions of this paper.

\section{Darboux transformation and $N$th-order rational solutions}
\label{Sec2}
In the frame of the Ablowitz-Kaup-Newell-Segur inverse scattering~\cite{AKNS}, Eq.~\eref{NNLS} has the Lax pair in the form:
\begin{subequations}
\label{CNLS4}
\begin{align}
& \Psi_x=U\Psi=\begin{pmatrix}
\lam & u \\
-\varepsilon \hat{u} & -\lam
\end{pmatrix}\Psi,\label{CNLS4a} \\
& \Psi_t=V\Psi=\begin{pmatrix}
-2 i \lam^2- i \varepsilon u \hat{u} & -2\, i \lam u - i u_x\\
2  i \varepsilon \lam   \hat{u} -  i \varepsilon \hat{u}_x &
2 i \lam^2 +  i \varepsilon u \hat{u}
\end{pmatrix}\Psi,\label{CNLS4b}
\end{align}
\end{subequations}
where $ \Psi=(f, g)^T$ (the superscript $T$
represents the vector transpose) is the vector eigenfunction, $
\lambda $ is the spectral parameter, the $hat$ denotes the combination of complex conjugate and space reversal, and Eq.~\eref{NNLS} can be
recovered from the compatibility condition $U_t - V_x + U\,V -
V\,U=0$. It is easy to prove that if $(f, g)^T$ is a nonzero solution of Lax pair~\eref{CNLS4}, then $(\hat{g}, \varepsilon \hat{f})^T$ also solves the same Lax pair in replacement of $\lam$ with $\lam^*$.
Based on the previous work in Ref.~\cite{LiXu}, we present the $N$-fold DT of Eq.~\eref{NNLS} as follows:
\begin{proposition}
\label{Prop21}
Assume that $\Psi_k=(f_k, g_k)^T$ ($ 1\leq k\leq N$) are a set of $N$ linearly-independent solutions of Eqs.~\eref{CNLS4} with different spectral parameters $\lam_k\in\mathbb{C}\setminus\mathbb{R}$. Then, the new eigenfunction $\Psi_{[N]}$ and new potential $u_{[N]}$
\begin{align}
& \Psi_{[N]}=T_{[N]}\Psi, \quad   T_{[N]}=
\begin{pmatrix}
\lam^N  - \sum\limits_{n=1}^{N}a_n(x,t) \lam^{n-1}  & - \sum\limits_{n=1}^{N}b_n(x,t) (-\lam)^{n-1}  \\
- \sum\limits_{n=1}^{N}c_n(x,t) \lam^{n-1}  & \lam^N  -
\sum\limits_{n=1}^{N}d_n(x,t)
(-\lam)^{n-1}
\end{pmatrix} , \label{ETR}\\[1.2mm]
&  u_{[N]} = u +2\,(-1)^{N-1} b_{N}, \quad \hat{u}_{[N]} =
\hat{u} + 2\,\varepsilon c_{N}, \label{PotentialTrana}
\end{align}
also solve the Lax pair~\eref{CNLS4}, where
$a_n$, $b_n$, $c_n$ and $d_n$ $( 1\leq  n \leq N )$ are uniquely determined by
\begin{align}
& T_{[N]}\big|_{\lam=\lam_k}\!
\begin{pmatrix}
f_k\\
g_k
\end{pmatrix}=\mb{0}, \quad
T_{[N]}\big|_{\lam=\lam^*_k}\!
\begin{pmatrix}
\hat{g}_k \\
\varepsilon  \hat{f}_k
\end{pmatrix} =\mb{0} \quad (1 \leq k \leq N).
\label{UndetCoeff}
\end{align}
Particularly, the functions $b_N$ and $c_N$ can be represented  as
\begin{align}
& b_N= (-1)^{N-1}\frac{\tau_{N+1,N-1}}{\tau_{N,N}},\quad c_N =
 \frac{\tau_{N-1,N+1}}{\tau_{N,N}},
\label{PotentialTranb}
\end{align}
with
\begin{align}
 \tau_{M, L}=
\begin{vmatrix}
F_{N\times M} & G_{N\times L} \\
\varepsilon \tilde{G}_{N\times M} & \tilde{F}_{N\times L}
\end{vmatrix} \quad (M+L=2N), \label{tao}
\end{align}
where the block matrices $ F_{N\times M} =
\big(\lam_k^{m-1}f_k\big)_{1\leqslant k \leqslant N, \atop 1\leqslant m \leqslant M}$,
$G_{N\times L}= \big((-\lam_k)^{m-1}g_k\big)_{1\leqslant k \leqslant N, \atop 1\leqslant m \leqslant L}$,
$\tilde{G}_{N\times M} =
\big(\lam^{*m-1}_k \hat{g}_k\big)_{1\leqslant k \leqslant N, \atop 1\leqslant m \leqslant M}$ and $\tilde{F}_{N\times L} =
\big((-\lam^*_k)^{m-1}\hat{f}_k\big)_{1\leqslant k \leqslant N, \atop 1\leqslant m \leqslant L}$.
\end{proposition}

By choosing the plane-wave solution of Eq.~\eref{NNLS} with $\varepsilon = -1$ as a seed
\begin{align}
u_{\rm pw}=\rho e^{2 i \rho ^2 t+i \phi }, \label{cws}
\end{align}
where $\rho \neq 0$ and $\phi$ are two real parameters, we can use the above DT to obtain the exponential soliton solutions~\cite{LiXu}. At $\lambda =\lambda_k $ ($\lambda_k \neq i\sigma \rho, \sigma=\pm1$), the solutions of Eqs.~\eref{CNLS4} with the potential given by~\eref{cws} are obtained as
\begin{align}
\begin{pmatrix}
f_k\\
g_k
\end{pmatrix}=
\begin{pmatrix}
 e^{i \rho^2 t + \frac{i}{2}\phi}\big(\alpha_k  e^{ \mu_k  \chi_k}+\beta_k  e^{ -\mu_k  \chi_k }\big)\\
 e^{-i \rho^2 t - \frac{i}{2}\phi}\big[\frac{\alpha_k(\mu_k-
\lam_k)}{\rho}  e^{\mu_k \chi_k } - \frac{\beta_k(\mu_k + \lam_k
)}{\rho}  e^{ -\mu_k \chi_k }\big]
\end{pmatrix} \quad (1 \leq k \leq N), \label{lpsol}
\end{align}
where $\mu_k =\sqrt{\lam_k^2 + \rho^2}$, $\chi_k =x- 2\, i  \lam_k
t$, $\alpha_k$ and $\beta_k $  are two nonzero constants in $\mathbb{C}$.
If $\lambda_1, \lambda_2,\ldots,\lambda_N \to i\sigma \rho$ ($\sigma=\pm1$),
solution~\eref{lpsol} reduces to a rational one, so that a chain of rational solutions can be derived in the limit.
In doing so, we assume that $\lam_{k} =  i \sigma \rho(1+\delta_k)$ ($1\leq k \leq N$) with $\delta_k$ being small parameter,
and also introduce $\delta_k$ into the constants $\alpha_k $ and $\beta_k $ in the form
\begin{align}
\alpha_k = e^{\mu_k \sum_{l=1}^{\infty} s_l \delta_k^{l-1}}, \quad
\beta_k = -e^{-\mu_k \sum_{l=1}^{\infty} s_l \delta_k^{l-1}} \quad (1\leq k \leq N),
\end{align}
where $s_l$'s are arbitrary constants in $\mathbb{C}$. Then, we expand all the elements of $\tau_{M,L}$ ($M+L=2N$) in the Taylor series of $\delta_k$ as follows:
\begin{equation}
\lam_k^{m-1}f_k=\sum^{\infty}_{j=1}f^{(j-1,m-1)}\delta_k^{j-1},\quad
\lam_k^{m-1}g_k=\sum^{\infty}_{j=1}g^{(j-1,m-1)}\delta_k^{j-1},  \label{Taylor}
\end{equation}
where $1\leq m \leq \max\{M,L\}$ and $1\leq k \leq N$.

Next, substituting Eq.~\eref{Taylor} into $\tau_{M,L}$ and taking the limit, we have the following result:
\begin{proposition}
\label{Prop22}
With the plane-wave solution~\eref{cws} as a seed and under the degeneracy $\lambda_1, \lambda_2,\ldots,\lambda_N \to i\sigma \rho$ ($\sigma=\pm1$), the potential transformation~\eref{PotentialTrana} can yield the $N$th-order rational solutions for Eq.~\eref{NNLS} with $\varepsilon = -1$:
\begin{align}
u_{[N]} =\rho e^{2 i \rho ^2 t+i \phi }+2\,\frac{\tau'_{N+1,N-1}}{\tau'_{N,N}}, \quad
\hat{u}_{[N]} = \rho e^{-2 i \rho ^2 t-i \phi } - 2\frac{\tau'_{N-1,N+1}}{\tau'_{N,N}},\label{NPT}
\end{align}
with \begin{align}
\tau'_{M, L}=
\begin{vmatrix}
F'_{N\times M} & G'_{N\times L} \\
-\tilde{G}'_{N\times M} & \tilde{F}'_{N\times L} \\
\end{vmatrix}\quad (M+L=2N),\label{PTR}
\end{align}
where the block matrices $F'_{N\times M}=\left(f^{(j-1,m-1)}\right)_{1\leqslant j \leqslant N,\atop 1\leqslant m \leqslant M}$, $G'_{N\times L}=\left((-1)^{m-1}g^{(j-1,m-1)}\right)_{1\leqslant j \leqslant N,\atop 1\leqslant m \leqslant L}$, $\tilde{G}'_{N\times M}=\left(\hat{g}^{(j-1,m-1)}\right)_{1\leqslant j \leqslant N,\atop 1\leqslant m \leqslant M}$ and
 $\tilde{F}'_{N\times L}=\Big((-1)^{m-1} \hat{f}^{(j-1,m-1)}\Big)_{1\leqslant j \leqslant N,\atop 1\leqslant m \leqslant L}$.
 \end{proposition}
\beginproof
Via the determinant properties, we expand $\tau_{M,L}$ in the following form
\begin{align}
\tau_{M,L}
=&\sum_{\substack{i_1,\dots, i_N \in \mathbb{Z}^+; \\
j_1, \dots, j_N \in \mathbb{Z}^+}} \prod_{k=1}^{N}\delta_k^{i_k-1}\delta_k^{*j_k-1}\notag\\
&\times\begin{vmatrix}
f^{(i_1-1,0)}&\dots&f^{(i_1-1,M-1)}&g^{(i_1-1,0)}&\dots &(-1)^{L-1}g^{(i_1-1,L-1)}\\
\vdots&\ddots&\vdots&\vdots&\ddots&\vdots\\
f^{(i_N-1,0)}&\dots&f^{(i_N-1,M-1)}&g^{(i_N-1,0)} &\dots&(-1)^{L-1}g^{(i_N-1,L-1)}\\
-\hat{g}^{(j_1-1,0)}&\dots&-\hat{g}^{(j_1-1,M-1)}&\hat{f}^{(j_1-1,0)}&\dots &(-1)^{L-1}\hat{f}^{(j_1-1,L-1)}\\
\vdots&\ddots&\vdots&\vdots&\ddots&\vdots\\
-\hat{g}^{(j_N-1,0)}&\dots&-\hat{g}^{(j_N-1,M-1)}&\hat{f}^{(j_N-1,0)}&\dots &(-1)^{L-1}\hat{f}^{(j_N-1,L-1)}
\end{vmatrix}\notag \\
\notag \\
=&  \Delta_N
\begin{vmatrix}
f^{(0,0)}&\dots&f^{(0,M-1)}&g^{(0,0)}&\dots &(-1)^{L-1}g^{(0,L-1)}\\
\vdots&\ddots&\vdots&\vdots&\ddots&\vdots\\
f^{(N-1,0)}&\dots&f^{(N-1,M-1)}&g^{(N-1,0)} &\dots&(-1)^{L-1}g^{(N-1,L-1)}\\
-\hat{g}^{(0,0)}&\dots&-\hat{g}^{(0,M-1)}&\hat{f}^{(0,0)}&\dots &(-1)^{L-1}\hat{f}^{(0,L-1)}\\
\vdots&\ddots&\vdots&\vdots&\ddots&\vdots\\
-\hat{g}^{(N-1,0)}&\dots&-\hat{g}^{(N-1,M-1)}&\hat{f}^{(N-1,0)}&\dots &(-1)^{L-1}\hat{f}^{(N-1,L-1)}
\end{vmatrix}
+ \text{small terms},
\label{tauML}
\end{align}
where $\Delta_N$ is defined by
\begin{align}
\Delta_N:=
\sum_{\substack{i_1, \dots, i_N\in [N]; \\ j_1, \dots, j_N\in [N]}}(-1)^{p_{i_1, \dots, i_N}+p_{j_1, \dots, j_N}}\prod_{k=1}^{N}\delta_k^{i_k-1} \delta_k^{*j_k-1},
\end{align}
with $i_1\neq i_2\neq\cdots\neq i_N$, $j_1\neq j_2\neq\cdots\neq j_N$, $[N]:=\{1,\dots, N\}$, and
\begin{align}
&p_{i_1, \dots, i_N}=\left\{
\begin{array}l
1,~~~~~(i_1, \dots, i_N) \text{~is an odd permutation},\\
0,~~~~~(i_1, \dots, i_N) \text{~is an even permutation},
\end{array}
\right. \notag  \\
&p_{j_1, \dots, j_N}=\left\{
\begin{array}l
1,~~~~~(j_1, \dots, j_N) \text{~is an odd permutation},\\
0,~~~~~(j_1, \dots, j_N) \text{~is an even permutation}.
\end{array}
\right. \notag
\end{align}
Then, plugging~\eref{tauML} into the  transformation~\eref{PotentialTrana} and calculating the limit at $\delta_k \to 0$ ($1\leq k \leq N$), we immediately arrive at solution~\eref{NPT}.   \hfill  \endproof

\begin{remark}
In principle, the expansion coefficients $f^{(j-1,m-1)} $ and $g^{(j-1,m-1)}$ in Eq.~\eref{Taylor} can be obtained from
\begin{align}
& f^{(j-1,m-1)} =\frac{1}{(j-1)!}\frac{\partial^{j-1}(\lam_{k}^{m-1} f_k)}{\partial\lam^{j-1}_{k}}\bigg|_{\delta_k=0}, \quad
g^{(j-1,m-1)} =\frac{1}{(j-1)!}\frac{\partial^{j-1}(\lam_{k}^{m-1} g_k)}{\partial\lam^{j-1}_{k}}\bigg|_{\delta_k=0}. \label{Taylor2}
\end{align}
With the aid of some symbolic computation softwares like Mathematica, one can calculate those coefficients in a recursive manner. In Appendix~\ref{appendixA}, we present the formulas for $f^{(j-1,m-1)} $ and $g^{(j-1,m-1)}$ ($1\leq j \leq 4$, $1\leq m \leq 5$), which will be used in obtaining the rational solutions with the order $1\leq N \leq 4$.
\end{remark}

\section{Asymptotic analysis of the rational solutions}
\label{Sec3}

In this section, we will study the asymptotic behavior of solution~\eref{NPT}
with $1\leq N \leq 4$ by deriving  all non-plane-wave asymptotic expressions as $|t|\to\infty$. 
Throughout this section, we use $\mathcal{C}$ to represent the center trajectories of asymptotic solitons,
and use the subscript $R$ and $I$ to respectively denote the real and imaginary parts of $s_1$.
Besides, we say that  there is no asymptotic soliton along $\mathcal{C}$ for solution~\eref{NPT}
if its asymptotic limit as $|t|\to\infty$ is just a plane wave.

\subsection{First-order rational solution}
\label{Sec3.1}

With $N=1$ in solution~\eref{NPT}, we have the first-order rational solution of Eq.~\eref{NNLS} with $\epsilon=-1$ as follows:
\begin{align}
u_{[1]} =\rho e^{2 i \rho^2 t+i \phi} \left[1-\frac{\left(2\rho\xi+K-i\sigma\right)\left(2\rho\eta-K^{*}+i\sigma\right)}
{2\rho^2\xi\eta+\rho K \eta-\rho K^{*} \xi-\frac{1}{2}\left(|K|^2+1\right)}\right],\label{1solution}
\end{align}
where $\xi=x+2\sigma\rho t$, $\eta=x-2\sigma\rho t$, $K=2\rho s_1+i\sigma$, and $\sigma=\pm 1$.
By separating the real and imaginary parts of the denominator of solution~\eref{1solution}, it can be found that this solution is nonsingular if and only if $s_{1I} \neq -\frac{\sigma}{2\rho}$. In Ref.~\cite{LiXu1}, we have obtained two asymptotic solitons for solution~\eref{1solution} by setting $x+2\sigma \rho t =\mathcal{O}(1) $ and $x-2\sigma \rho t = \mathcal{O}(1)  $ as $|t|\ra\infty$, respectively. Next, we present a rigorous proof that $u_{[1]}$ has only two asymptotic solitons, which are respectively along the curves $x \mp 2\sigma\rho t \mp s_{1R}=0 $.

Observing that $u_{[1]}$ is explicitly dependent on $\xi$ and $\eta$ (without consideration of the phase part $e^{2 i \rho^2 t+i \phi}$), we first give the asymptotic behavior of $\xi$ and $\eta$ as $|t|\to\infty$ when the asymptotic solitons are formed.
\begin{proposition}
\label{prop0}
As $|t|\ra\infty$,  the asymptotic solitons of solution~\eref{1solution} are formed only when $\xi$ and $\eta$ satisfy the asymptotic relation:
\begin{align}
(i)\, |\xi|=\mathcal{O}(|t|),\, |\eta|=\mathcal{O}(1),\,\,\, \text{or}\,\,\, (ii)\, |\eta|=\mathcal{O}(|t|),\, |\xi|=\mathcal{O}(1).  \label{AsyRel1}
\end{align}
\end{proposition}

\beginproof  Because $\xi-\eta=4\sigma \rho t$, there are three possibilities for the asymptotic behavior of $\xi$ and $\eta$: (a) $|\xi|=\mathcal{O}(|t|),\, |\eta|=\mathcal{O}(|t|^\alpha)$ ($0\leq \alpha<1$); (b) $|\eta|=\mathcal{O}(|t|),\, |\xi|=\mathcal{O}(|t|^\beta)$ ($0\leq \beta<1$); (c) $\mathcal{O}(|\xi|) = \mathcal{O}(|\eta|) \gg \mathcal{O}(|t|)$. For cases (a) and (b) with $\alpha=\beta=0$, $u_{[1]}$ goes to two non-plane-wave asymptotic states as $|t|\to \infty$; whereas for the other cases  the asymptotic limits of  solution~\eref{1solution} are always $-\rho e^{2 i \rho^2 t+i \phi}$. Therefore, the  asymptotic solitons of $u_{[1]}$ appear only when  the asymptotic relation in~\eref{AsyRel1} is satisfied. \hfill \endproof

Based on Proposition~\ref{prop0}, we can explicitly determine the asymptotic solitons of $u_{[1]}$ as well as their center trajectories.

\begin{theorem}
\label{Thm1}
Asymptotically as $|t| \to \infty $, the first-order rational solution~\eref{1solution} admits two asymptotic soliton states:
\begin{subequations}
\begin{align}
& u^{(1)}_{[1]}=\rho e^{2 i \rho^2 t+i \phi} \left[1-\frac{4\rho W_1(x,t) -4 \rho s_1^*
+ 4 i\sigma}{2\rho W_1(x,t)- 2\rho s^*_1  + i\sigma}\right], \quad W_1(x,t)=\eta, \label{1asy1a}\\[1.5mm]
&
u^{(2)}_{[1]}= \rho e^{2 i \rho^2 t+i \phi} \left[1-\frac{4\rho W_2(x,t) + 4 \rho s_1}{2\rho W_2(x,t) + 2\rho s_1 + i\sigma }\right],
\quad\,\,\, W_2(x,t)=\xi,  \label{1asy2a}
\end{align}
\end{subequations}
whose center trajectories are respectively given by the straight lines
\begin{align}
& \mathcal{C}^{(1)}_{[1]}:\, x-2\sigma\rho t = s_{1R},  \quad
\mathcal{C}^{(2)}_{[1]}:\,   x+2\sigma\rho t=-s_{1R}. \label{Traj1}
\end{align}
Moreover, $u_{[1]}$ approaches the asymptotic solitons $u^{(1,2)}_{[1]}$ respectively  along the curves $\mathcal{C}^{(1,2)}_{[1]}$ at the  rate of $\mathcal{O}(t^{-1})$.
\end{theorem}

\beginproof
For cases (i) and (ii) in~\eref{AsyRel1}, we rewrite solution~\eref{1solution} respectively in the following equivalent forms
\begin{align}
& u_{[1]} = \rho e^{2 i \rho^2 t+i \phi} \left[1-\frac{\left(2\rho\eta-K^*+i\sigma\right)\left(2\rho+(K-i\sigma)\xi^{-1}\right)}{2\rho^2\eta-\rho K^* + \rho K \eta \xi^{-1} - \frac{|K|^2 +1}{2}\xi^{-1}}\right], \label{1asy1b}  \\
& u_{[1]} = \rho e^{2 i \rho^2 t+i \phi} \left[1-\frac{\left(2\rho\xi+K-i\sigma\right)\left(2\rho+ (i\sigma-K^*)\eta^{-1}\right)}{2\rho^2\xi+\rho K -\rho K^{*}\xi\eta^{-1} - \frac{|K|^2+1}{2}\eta^{-1}}\right]. \label{1asy2b}
\end{align}
Then, we calculate the Taylor series expansion in terms of  $\xi^{-1}$ and $\eta^{-1}$ respectively for Eqs.~\eref{1asy1b} and~\eref{1asy2b}, obtaining that
\begin{align}
& u_{[1]} \sim \rho e^{2 i \rho^2 t+i \phi}\left[1-\frac{4\rho(\eta+s_1^*)-4K^*}{2\rho\eta-K^*}+\mathcal{O}\left(t^{-1}\right)\right], \label{1asy1c}  \\
& u_{[1]} \sim \rho e^{2 i \rho^2 t+i \phi}\left[1-\frac{4\rho(\xi+s_1)}{2\rho\xi+K}+\mathcal{O}\left(t^{-1}\right)\right], \label{2asy2c}
\end{align}
where $\mathcal{O}(\xi^{-1})$ and $\mathcal{O}(\eta^{-1})$  have been replaced by $\mathcal{O}(t^{-1})$ through~\eref{AsyRel1}. As a result,  $u_{[1]}$ goes to two asymptotic solitons $u^{(1)}_{[1]}$ and $u^{(2)}_{[1]}$ (which are associated with cases (i) and (ii) in~\eref{AsyRel1}, respectively) at the rate of $\mathcal{O}(t^{-1})$. Meanwhile, by the extreme value analysis of $\big|u^{(1,2)}_{[1]}\big|^2$, we find that they have a unique extremum respectively at $W_1=s_{1R}$ and $W_2=-s_{1R}$, which  determine the center trajectories $ \mathcal{C}^{\pm}_{1,2}$ given in~\eref{Traj1}.
\hfill   \endproof

\subsection{Second-order rational solution}
\label{Sec3.2}
By taking $N=2$ in solution~\eref{NPT}, we present the second-order rational solution of Eq.~\eref{NNLS} with $\epsilon=-1$ as follows:
\begin{align}
u_{[2]} &=\text{$\rho $e}^{2 i \rho ^2 t+i \phi }+
\frac{8i \sigma\text{$\rho^2 $e}^{2 i \rho ^2 t+i \phi } \left( \rho ^2 \xi ^{\prime 2} R+\rho^2\eta ^{\prime 2} S  +2 i \sigma \rho  \eta^{\prime }  S   - S  \right)}
{ \rho ^2 (\xi ^{\prime 2} P +\eta ^{\prime 2}  Q+2\xi^{\prime }\eta^{\prime })+ 2 \rho  \xi^{\prime }  (2 \rho  R+i\sigma )+2 \rho \eta^{\prime } (2  \rho S +i \sigma )-4 \rho ^2 R S -1}, \label{2solution}
\end{align}
with
\begin{align}
&P=1-4 i \sigma  \rho   R,\quad  R=-\frac{2}{3} \rho ^2 \eta ^{\prime 3} +2 \sigma  \rho   t-2 i \sigma  \rho \eta ^{\prime 2} +\eta^{\prime } +s_2^*, \notag \\
&Q=1-4 i  \sigma \rho  S,\quad S=\frac{2}{3} \rho ^2 \xi ^{\prime 3} +2 \sigma \rho   t+\xi^{\prime } +s_2, \notag\\
&\xi^{\prime }=\xi+s_1,\quad \eta^{\prime }=\eta-s_1^*, \quad \xi=x+2\sigma\rho t, \quad \eta=x-2\sigma\rho t. \notag
\end{align}

By neglecting the phase part $e^{2 i \rho^2 t+i \phi}$, it can be regarded that solution~\eref{2solution} is explicitly dependent on $\xi$, $\eta$ and $t$. Thus,  we need to consider the asymptotic relations among $\xi$, $\eta$ and $t$ when the asymptotic solitons appear as $|t| \to \infty$.

\begin{lemma}\label{lemma2}
Along the soliton center trajectories $\mathcal{C}$ as $|t| \ra \infty$, $\xi$ and $\eta$ in solution~\eref{2solution} obey the following asymptotic properties:
\begin{enumerate}
\item[(i)]
$|\xi|, |\eta| \ra \infty$ as $|t| \ra \infty$, in other words, $|\xi|, |\eta| \neq \mathcal{O}(1)$ or $0$;

\item[(ii)]

$|\xi|$ and $|\eta|$ cannot be of the same order, i.e., $|\xi| \neq \mathcal{O}(|\eta|)$.
\end{enumerate}
\end{lemma}

\beginproof  (i) There are three different cases when $|\xi|$ or $|\eta| \not\to \infty$.  First, if $|\xi|,|\eta|=\mathcal{O}(1)$ or $0$, then $\xi-\eta=4\sigma \rho t=\mathcal{O}(1)$ or $0$, which contradicts with $|t|\ra \infty$. Second, if $|\xi|=\mathcal{O}(1)$ or $0$ but $|\eta| \ra\infty$, then $|\eta| = |\xi-4\sigma \rho t| = \mathcal{O}(|t|)$. Thus, solution~\eref{2solution} is dominated by
\begin{align}
u_{[2]} &\sim  e^{2 i \rho^2 t+i \phi}\left(\rho+\frac{3i\sigma t-i\rho \xi ^{\prime2} \eta}{\eta t}\right) . \label{A.1}
\end{align}
Third, if $|\eta|=O(1)$ or $0$ but $|\xi| = \mathcal{O}(|t|) $, 
we have
\begin{align}
u_{[2]} &\sim e^{2 i \rho^2 t+i \phi}\left(\rho-\frac{3i\sigma t+i\rho \eta ^{\prime2} \xi}{\xi t}\right). \label{A.2}
\end{align}
As $|t|\to \infty$, both the right-hand sides of~\eref{A.1} and~\eref{A.2} go to $\rho e^{2 i \rho^2 t+i \phi}$, implying that no asymptotic soliton exists. Therefore, we know that $|\xi|, |\eta| \neq \mathcal{O}(1)$ or 0 along the curve $\mathcal{C}$ as $|t| \ra\infty$.

ii) With reduction to absurdity, we assume that $|\xi|=\mathcal{O}(|\eta|)$ along $\mathcal{C}$ as $|t|\ra\infty$. Recalling that $\xi=\eta+4\sigma\rho t$, there must be $|\xi|, |\eta| \gg  \mathcal{O}(|t|)$ or $|\xi|=\mathcal{O}(|\eta|)=\mathcal{O}(|t|)$. For both the two cases, solution~\eref{2solution} is dominated by
\begin{align}
u_{[2]} &\sim \rho e^{2 i \rho^2 t+i \phi} \left(1+12i\sigma\rho\frac{\eta-\xi}{\xi\eta}\right). \label{A.3}
\end{align}
Again, the right-hand side of~\eref{A.3} tends to $\rho e^{2 i \rho^2 t+i \phi}$, which implies that no asymptotic soliton appears. Hence, we obtain that $|\xi| \neq \mathcal{O}(|\eta|)$ along the curve $\mathcal{C}$ as $|t| \ra\infty$.   \hfill \endproof

\begin{remark}\label{remark2a}
The asymptotic solitons of solution~\eref{2solution} cannot be located in any straight line
$\mathcal{L}\!: x-ct=\text{const}$. 
Otherwise, because $\xi=(x- c\,t) + ( c+ 2 \sigma \rho )t$ and $\eta =(x- c\,t) + (c- 2 \sigma \rho )t$,
the asymptotic behavior of $\xi$ and $\eta$ along $\mathcal{L}$ can be obtained by
\begin{align}\left\{
\begin{array}{l}
|\xi|=\mathcal{O}(|t|),\, |\eta|=\mathcal{O}(1),~~~~~~c=2\sigma\rho, \\
|\xi|=\mathcal{O}(1),\,\, |\eta|=\mathcal{O}(|t|),~~~~~c=-2\sigma\rho, \\
|\xi|=\mathcal{O}(|t|),\, |\eta|=\mathcal{O}(|t|),~~~~c\neq\pm2\sigma\rho,
\end{array}
\right.
\end{align}
all of which violate the properties in Lemma~\ref{lemma2}. As a result,  the asymptotic limits of solution~\eref{2solution} always produce a plane wave, as discussed in Eqs.~\eref{A.1}--\eref{A.3}. That is to say, the asymptotic solitons of solution~\eref{2solution} should only be localized in some curves in the $xt$ plane.
\end{remark}

\begin{proposition}
\label{prop1}
As $|t|\ra\infty$,  the asymptotic solitons of solution~\eref{2solution} are formed only when $\xi$, $\eta$ and $t$ satisfy the asymptotic relation:
\begin{align}
(i)\, |\xi|=\mathcal{O}(|t|), \,\, t\eta^{-2} - \frac{\sigma\rho}{3}\eta=\mathcal{O}(1),\,\,\, \text{or} \,\,\, (ii)\, |\eta|=\mathcal{O}(|t|), \, t\xi^{-2} + \frac{\sigma\rho}{3}\xi =\mathcal{O}(1).  \label{AsyRel2}
\end{align}
\end{proposition}

\beginproof
Note from Lemma~\ref{lemma2} that $|\xi|, |\eta| \neq \mathcal{O}(1)$ or $0$ and $|\xi| \neq \mathcal{O}(|\eta|)$ along the soliton center trajectories $\mathcal{C}$, and recall that
$\xi=\eta+4\sigma\rho t$ always holds true. Thus, there are only two possibilities for the asymptotic behavior of $\xi$ and $\eta$ along $\mathcal{C}$ as $|t|\ra\infty$: (a) $|\xi|=\mathcal{O}(|t|)$, $|t|=\mathcal{O}(|\eta|^\alpha)$ $(\alpha >1)$;  (b) $|\eta|=\mathcal{O}(|t|)$, $|t|=\mathcal{O}(|\xi|^\beta)$ $(\beta >1)$. For such two cases, extracting the most possible dominant terms in solution~\eref{2solution} gives
\begin{align}
& u_{[2]} \sim \rho  e^{2 i \rho^2 t+i \phi}\left[1 +\frac{2i\sigma \eta^2}{\frac{2}{3}\rho\eta^3+(i\sigma-2\rho s_1^*)\eta^2-2\sigma t}\right], \quad |\xi|=\mathcal{O}(|t|),\, |t|=\mathcal{O}(|\eta|^\alpha), \label{2medisolutiona} \\
& u_{[2]} \sim \rho  e^{2 i \rho^2 t+i \phi}\left[1 - \frac{2i\sigma \xi^2}{\frac{2}{3}\rho\xi^3+(i\sigma+2\rho s_1)\xi^2+2\sigma t}\right], \quad |\eta|=\mathcal{O}(|t|),\, |t|=\mathcal{O}(|\xi|^\beta).  \label{2medisolutionb}
\end{align}

Then, considering the balance between $t$ and $\eta$ or between $t$ and $\xi$, we have
\begin{align}
\label{2balance-1}
u_{[2]} \sim \left\{
\begin{array}{l}
\rho e^{2i\rho^2 t+i \phi }\left(1 +\frac{3i\sigma}{\rho \eta}\right),~~~~~~~~~~~~~~~~~~~~~~~~~~|\xi|=\mathcal{O}(|t|),\, \mathcal{O}(|\eta|)\ll |t|\ll \mathcal{O}(|\eta|^3),\\[1mm]
\rho e^{2i\rho^2 t+i \phi }\left(1 -\frac{i \eta^2}{t}\right),~~~~~~~~~~~~~~~~~~~~~~~~~~|\xi|=\mathcal{O}(|t|),\, |t|\gg \mathcal{O}(|\eta|^3),\\[1mm]
\rho e^{2i\rho^2 t+i \phi }\left(1 +\frac{3i\sigma \eta^2}{\rho\eta^3-3\sigma t}\right),~~~~~~~~~~~~~~~~~~|\xi|=\mathcal{O}(|t|),\, t\sim V(x,t) \eta^3\,\,\,(V \neq\frac{\sigma\rho}{3}),\\[1mm]
\rho e^{2i\rho^2 t+i \phi }\left(1-\frac{i\eta^2}{W(x,t)\eta^\gamma}\right),~~~~~~~~~~~~~~~~~|\xi|=\mathcal{O}(|t|),\, t\sim \frac{\sigma\rho}{3}\eta^3+W(x,t)\eta^\gamma\,\,\,(2<\gamma<3),\\[1mm]
\rho e^{2i\rho^2 t+i \phi }\left(1 +\frac{2i\sigma}{i\sigma-2\rho s_1^*}\right), ~~~~~~~~~~~~~~~~~~~|\xi|=\mathcal{O}(|t|),\, t\sim \frac{\sigma\rho}{3}\eta^3+W(x,t)\eta^\gamma\,\,\,(\gamma<2),\\[1mm]
\rho e^{2i\rho^2 t+i \phi }\left(1 -\frac{2i\sigma}{2\sigma W(x,t)+2\rho s_1^*-i\sigma}\right),~~~|\xi|=\mathcal{O}(|t|),\, t\sim \frac{\sigma\rho}{3}\eta^3+W(x,t)\eta^2,
\end{array}
\right.
\end{align}
and
\begin{align}
\label{2balance-2}
u_{[2]} \sim \left\{
\begin{array}l
\rho e^{2 i \rho ^2 t+i \phi }\left(1-\frac{3i\sigma}{\rho \xi}\right),~~~~~~~~~~~~~~~~~~~~~~~~~|\eta|=\mathcal{O}(|t|),\,  \mathcal{O}(|\xi|) \ll t \ll \mathcal{O}(|\xi|^3), \\[1mm]
\rho e^{2 i \rho^2 t+i \phi }\left(1-\frac{i\xi^2}{t}\right),~~~~~~~~~~~~~~~~~~~~~~~~~~|\eta|=\mathcal{O}(|t|),\, |t|\gg \mathcal{O}(|\xi|^3),  \\[1mm]
\rho e^{2 i \rho ^2 t+i \phi }\left(1 -\frac{3i\sigma \xi^2}{\rho\xi^3+3\sigma t}\right),~~~~~~~~~~~~~~~~~~|\eta|=\mathcal{O}(|t|),\,t\sim V(x,t) \xi^3 \,\, (V \neq-\frac{\sigma\rho}{3}),     \\[1mm]
\rho e^{2 i \rho ^2 t+i \phi }\left(1 - \frac{i \xi^2}{W(x,t)\xi^\delta}\right),~~~~~~~~~~~~~~~~~|\eta|=\mathcal{O}(|t|)\, ,t\sim -\frac{\sigma\rho}{3}\xi^3+W(x,t)\xi^\delta \,\, (2<\delta<3),  \\[1mm]
\rho e^{2 i \rho ^2 t+i \phi }\left(1-\frac{2i\sigma}{i\sigma+2\rho s_1}\right),~~~~~~~~~~~~~~~~~~~|\eta|=\mathcal{O}(|t|),\,t\sim -\frac{\sigma\rho}{3}\xi^3+W(x,t)\xi^\delta\,\, (\delta<2),   \\[1mm]
\rho e^{2 i \rho ^2 t+i \phi }\left(1-\frac{2i\sigma}{2\sigma W(x,t) + 2\rho s_1 + i\sigma}\right),~~~|\eta|=\mathcal{O}(|t|),\, t\sim -\frac{\sigma\rho}{3}\xi^3+W(x,t)\xi^2,
\end{array}
\right.
\end{align}
where $V(x,t), W(x,t)=\mathcal{O}(1)$. Apparently, the asymptotic limits of $u_{[2]}$ are a plane wave except for the last cases in Eqs.~\eref{2balance-1} and~\eref{2balance-2}, which respectively correspond to the two asymptotic relations in~\eref{AsyRel2}.
\hfill \endproof

With the availability of Proposition~\ref{prop1}, one can immediately obtain the explicit expressions of asymptotic solitons of solution~\eref{2solution} and determine their center trajectories at the same time.

\begin{theorem} \label{Thm2}
Asymptotically as $|t| \to \infty $, the second-order rational solution~\eref{2solution} admits two asymptotic soliton states:
\begin{subequations}
\label{2asy1}
\begin{align}
u^{(1)}_{[2]} = \rho e^{2i\rho^2 t+i \phi }\left(1 - \frac{2i\sigma}{2\sigma W_1(x,t)+2\rho s_1^* -i\sigma}\right), \label{2asy1a}\\[1.5mm]
u^{(2)}_{[2]} = \rho e^{2 i \rho ^2 t+i \phi }\left(1-\frac{2i\sigma}{2\sigma W_2(x,t)+2\rho s_1 +i\sigma}\right), \label{2asy2a}
\end{align}
\end{subequations}
with
\begin{align}
W_1=t\eta^{-2}-\frac{\sigma\rho}{3}\eta, \quad  W_2=t\xi^{-2}+\frac{\sigma\rho}{3}\xi.  \label{2W12}
\end{align}
Moreover,  the center trajectories of asymptotic solitons $u^{(1,2)}_{[2]}$ are given by the algebraic curves
\begin{align}
& \mathcal{C}^{(1)}_{[2]}:\,  t=\frac{\sigma\rho}{3}\left(\eta^3- 3s_{1R} \eta^2\right),  \quad
\mathcal{C}^{(2)}_{[2]}:\,   t=-\frac{\sigma\rho}{3}\left(\xi^3 + 3 s_{1R}\xi^2\right), \label{Traj2}
\end{align}
and $u_{[2]}$ approaches the asymptotic solitons $u^{(1,2)}_{[2]}$ respectively along the curves $\mathcal{C}^{(1,2)}_{[2]}$ at the rate of $\mathcal{O}(t^{-1/3})$.
\end{theorem}

\beginproof  Associated with the cases (i) and (ii) in~\eref{AsyRel2}, solution~\eref{2solution} can be respectively written in the following forms
\begin{align}
& u_{[2]}= e^{2 i \rho ^2 t+i \phi }\left[\rho + \frac{2\rho + 4(i\sigma-\rho s_1^*) \eta^{-1} +\mathcal{O}(\eta^{-2})}{1+2i\sigma \rho s_1^* + 2iW_1(x,t) -(2s_1^*+2i\sigma\rho s_1^{*2}) \eta^{-1} +\mathcal{O}(\eta^{-2}) }\right], \label{2asy1b} \\
& u_{[2]} = e^{2 i \rho ^2 t+i \phi }\left[\rho - \frac{2\rho+4\rho s_1\xi^{-1} +\mathcal{O}(\xi^{-2})}{1-2i\sigma\rho s_1-2i W_2(x,t) + (4 s_1-4i\sigma\rho s_1^2) \xi^{-1} +\mathcal{O}(\xi^{-2}) }\right], \label{2asy2b}
\end{align}
where the first equation has been obtained by dividing the numerator and denominator of~\eref{2solution} simultaneously by $\xi^3\eta^2$ and replacing $t$ via $t\sim \frac{\sigma\rho}{3}\eta^3+W_1(x,t)\eta^2$ ($W_1=\mathcal{O}(1)$), the second one has been obtained by dividing the numerator and denominator of~\eref{2solution} simultaneously by $\xi^2\eta^3$ and replacing $t$ via $t\sim -\frac{\sigma\rho}{3}\xi^3+W_2(x,t)\xi^2$ ($W_2=\mathcal{O}(1)$).

Then, we perform the Taylor series expansion in terms of $\eta^{-1}$ and $\xi^{-1}$  respectively for Eqs.~\eref{2asy1b} and~\eref{2asy2b}, obtaining that
\begin{align}
& u_{[2]}  \sim \rho e^{2 i \rho ^2 t+i \phi }\left[1 - \frac{2i\sigma}{2\sigma W_1(x,t)+2\rho s_1^* - i\sigma }+\mathcal{O}(t^{-1/3}) \right], \label{2asy1c}  \\
& u_{[2]} \sim  \rho e^{2 i \rho ^2 t+i \phi }\left[1 -\frac{2i\sigma}{2\sigma W_2(x,t) +2\rho s_1 + i\sigma}+\mathcal{O}(t^{-1/3})\right], \label{2asy2c}
\end{align}
where $\mathcal{O}(\eta^{-1})$ and $\mathcal{O}(\xi^{-1})$ have been replaced by $\mathcal{O}(t^{-\frac{1}{3}})$ through~\eref{AsyRel2} respectively for Eqs.~\eref{2asy1c} and~\eref{2asy2c}, and $W_{1,2}$ are defined in Eq.~\eref{2W12}. Therefore,  $u_{[2]}$ goes to the asymptotic solitons $u^{(1)}_{[2]}$ and $u^{(2)}_{[2]}$ (which are associated with the cases (i) and (ii) in~\eref{AsyRel2}, respectively) at the rate of $\mathcal{O}(t^{-1/3})$. Meanwhile,  calculating the derivative of $\big|u^{(i)}_{[2]}\big|^2$ with respect to $W_i$ ($i=1,2$) shows that $\big|u^{(i)}_{[2]}\big|^2$ has a unique extremum at $W_i=-\sigma \rho s_{1R}$ ($i=1,2$), which define the center trajectories $\mathcal{C}^{(1,2)}_{[2]}$ given in~\eref{Traj2}.
\hfill  \endproof

\subsection{Third-order rational solution}
\label{Sec3.3}

With $N=3$ in solution~\eref{NPT}, we have the third-order rational solution $u_{[3]}$, whose long expression is omitted here for brevity). Similarly to section~\ref{Sec3.2}, we perform an asymptotic analysis of the third-order rational solution.

It can be readily shown that $\xi $ and $\eta $ in $u_{[3]}$ enjoy the same properties as given in  Lemma~\ref{lemma2} and Remark~\ref{remark2a}. Thus, along the soliton center trajectories $\mathcal{C}$ as $|t|\ra\infty$, there are
just two admissible cases for the asymptotic behavior of $\xi$ and $\eta$: (a) $|\xi|=\mathcal{O}(|t|)$, $|t|=\mathcal{O}(|\eta|^\alpha)$ $(\alpha >1)$;  (b) $|\eta|=\mathcal{O}(|t|)$, $|t|=\mathcal{O}(|\xi|^\beta)$ $(\beta >1)$.
In what follows, we determine such asymptotic relations up to the subdominant level.

As for case (a),  we write the third-order rational solution in the form
\begin{align}
& u_{[3]} = \rho e^{2 i \rho^2 t+i \phi}\left[1
-\frac{2(\rho^2 \xi^6\eta^6 - 15\sigma\rho \xi^6 \eta^3 t-45\xi^6 t^2)
- 2(\rho s_1^*-i\sigma) (6\rho \xi^6\eta^5 -45\sigma \xi^6\eta^2 t)+ \cdots}
{(\rho^2 \xi^6 \eta^6 -15\sigma\rho \xi^6 \eta^3 t - 45\xi^6 t^2)
-(2\rho s_1^*-i\sigma)(3\rho\xi^6\eta^5 -\frac{45}{2}\sigma \xi^6\eta^2 t)
+ \cdots}\right], \label{3medisolutiona}
\end{align}
where $\rho^2 \xi^6\eta^6$, $- 15\sigma\rho \xi^6 \eta^3 t$ and  $-45\xi^6 t^2$ are the three possible most dominant
terms both in the numerator and denominator, and the dots stands for all the lower-power terms of $\xi$, $\eta$ and $t$.
It can be found that the asymptotic limit of $u_{[3]}$ yields the plane wave $-\rho e^{2 i \rho^2 t+i \phi}$ once $t \not\sim  \Big(\!\pm\!\frac{\sqrt{5}}{10}-\frac{\sigma}{6}\Big)\rho \eta^3 $. But with
$t \sim  \Big(\!\pm\!\frac{\sqrt{5}}{10}-\frac{\sigma}{6}\Big)\rho \eta^3 $, the most dominant contribution in
$\rho^2 \xi^6\eta^6 - 15\sigma\rho \xi^6 \eta^3 t -45 \xi^6 t^2$ vanish and those subdominant terms become the leading ones.
Then, we assume the asymptotic relation between $t$ and $\eta$ in a more specific form 
\begin{align}
t\sim   \bigg(\!\pm\!\frac{\sqrt{5}}{10}-\frac{\sigma}{6}\bigg)\rho \eta^{3} + W(x,t)\eta^{\gamma},  \label{AsyRel31}
\end{align}
where $0<\gamma<3$ is a real constant to be determined, and $ W(x,t)=\mathcal{O}(1)$. Keeping in mind the relation~\eref{AsyRel31} and $|\xi|=\mathcal{O}(|t|)$, we obtain the asymptotic limits of $u_{[3]}$ for different $\gamma$ as $|t|\to \infty$:
\begin{align}
& u_{[3]} \sim \left\{
\begin{array}{c}
-\rho e^{2i\rho^2 t+i \phi },~~~~~~~~~~~~~~~~~~~~~~~~~~~~~~~~~~~~~~~~~~~~~~~
t\sim \left(\pm\frac{\sqrt{5}}{10}-\frac{\sigma}{6}\right)\rho\eta^3+W\eta^\gamma\,\,\,(\gamma>2),\\[2mm]
\rho e^{2i\rho^2 t+i \phi }\left[1 -\frac{4(\rho s_1^*- i\sigma)}{2\rho s_1^*- i\sigma}\right],~~~~~~~~~~~~~~~~~~~~~~
t\sim \left(\pm\frac{\sqrt{5}}{10}-\frac{\sigma}{6}\right)\rho\eta^3+W\eta^\gamma\,\,\,(\gamma<2),\\[2mm]
\rho e^{2i\rho^2 t+i \phi }\left[1-\frac{8\sqrt{5}W(x,t) + 4(\pm 3-\sqrt{5}\sigma)(\rho s_1^*-i\sigma)}{4\sqrt{5}W(x,t)
+ (\pm 3-\sqrt{5}\sigma)(2\rho s_1^*-i\sigma)}\right],\quad  t\sim \Big(\!\pm\!\frac{\sqrt{5}}{10}-\frac{\sigma}{6}\Big)\rho \eta^3+W\eta^2,
\end{array}  \right.  \label{3balancec}
\end{align}
where only the last case describes a non-plane-wave asymptotic state, which implies that there must be $\gamma=2$ in~\eref{AsyRel31} in obtaining the asymptotic solitons of $u_{[3]}$.

Similarly for case (b), we write the third-order rational solution in another form
\begin{align}
& u_{[3]} = \rho e^{2 i \rho^2 t+i \phi}\left[1-\frac{2(\rho^2 \eta^6 \xi^6  + 15\sigma\rho\eta^6 \xi^3 t - 45\eta^6 t^2) + 2\rho s_1(6\rho \eta^6 \xi^5 + 45\sigma \eta^6 \xi^2 t ) + \cdots}{(\rho^2\eta^6 \xi^6
+ 15\sigma\rho \eta^6 \xi^3 t - 45 \eta^6 t^2)
+ (2\rho s_1+i\sigma)(3\rho \eta^6\xi^5 +\frac{45}{2}\sigma \eta^6 \xi^2 t) + \cdots }\right], \label{3medisolutionb}
\end{align}
where $\rho^2 \eta^6 \xi^6$, $15\sigma\rho \eta^6 \xi^3 t$ and $- 45\eta^6 t^2$ are the three possible most dominant
terms both in the numerator and denominator, and the dots stands for all the lower-power terms of $\xi$, $\eta$ and $t$. Also, one can find that the asymptotic limit of $u_{[3]}$ is not
a plane wave unless $t \sim -\Big(\!\pm\!\frac{\sqrt{5}}{10}-\frac{\sigma}{6}\Big)\rho\xi^3$, under which the most dominant contribution in $\rho^2 \eta^6 \xi^6  + 15\sigma\rho  \eta^6 \xi^3 t - 45\eta^6 t^2 $ is removed. More specifically,
we assume the asymptotic relation between $t$ and $\xi$ in the form
\begin{align}
t\sim -\bigg(\!\pm\!\frac{\sqrt{5}}{10}-\frac{\sigma}{6}\bigg)\rho\xi^3 + W(x,t)\xi^{\delta}  \label{AsyRel32}
\end{align}
with $0<\delta <3$ and $ W(x,t)= \mathcal{O}(1)$, and obtain the asymptotic limits of $u_{[3]}$ for different $\delta$ as follows:
\begin{align}
u \sim \left\{
\begin{array}{l}
-\rho e^{2i\rho^2 t+i \phi },~~~~~~~~~~~~~~~~~~~~~~~~~~~~~~~~~~~~~~~~~
t\sim -\Big(\!\pm\!\frac{\sqrt{5}}{10}-\frac{\sigma}{6}\Big)\rho\xi^3 + W\xi^{\delta}\,\,\,(2<\delta<3),   \\[2mm]
\rho e^{2i\rho^2 t+i \phi }\left(1 -\frac{4\rho s_1}{2\rho s_1+ i\sigma}\right),~~~~~~~~~~~~~~~~~~~~~~~~~~
t\sim -\Big(\!\pm\!\frac{\sqrt{5}}{10}-\frac{\sigma}{6}\Big)\rho\xi^3 + W\xi^{\delta}\,\,\,(\delta<2),\\[2mm]
\rho e^{2i\rho^2 t+i \phi}\left[1-\frac{8\sqrt{5}W(x,t) + 4\left(\pm 3-\sqrt{5}\sigma\right)\rho s_1}{4\sqrt{5}W(x,t)
+\left(\pm3-\sqrt{5}\sigma\right)(2\rho s_1+i\sigma)}\right], \quad
t\sim -\Big(\!\pm\!\frac{\sqrt{5}}{10}-\frac{\sigma}{6}\Big)\rho\xi^3 + W\xi^{2}.
\end{array}\label{3balanced}
\right.
\end{align}
Again, only the last case in~\eref{3balanced} represents a non-plane-wave asymptotic state, which means that one must take $\delta=2$ in~\eref{AsyRel32} to obtain the asymptotic solitons of $u_{[3]}$.

To this stage, we obtain the asymptotic relations that $\xi$, $\eta$ and $t$ must satisfy when the asymptotic solitons of $u_{[3]}$ appear, as stated in the following proposition:

\begin{proposition}
\label{prop2}
As $|t|\ra\infty$,  the asymptotic solitons of the third-order rational solution $u_{[3]}$ are formed only when $\xi$, $\eta$ and $t$ satisfy the asymptotic relation:
\begin{equation}
\label{AsyRel3}
\begin{aligned}
& (i)\,\, |\xi|=\mathcal{O}(|t|), \, t\eta^{-2}- \left(\!\pm\!\frac{\sqrt{5}}{10}-\frac{\sigma}{6}\right)\rho\eta=\mathcal{O}(1), \\
 \text{or}\,\,  & (ii)\,\, |\eta|=\mathcal{O}(|t|), \, t\xi^{-2}+\left(\!\pm\! \frac{\sqrt{5}}{10}-\frac{\sigma}{6}\right)\rho\xi=\mathcal{O}(1).
\end{aligned}
\end{equation}
\end{proposition}

Furthermore, we can derive the explicit expressions for all asymptotic solitons of $ u_{[3]}$ as well as their center trajectories.

\begin{theorem} \label{Thm3}
Asymptotically as $|t| \to \infty $, the third-order rational solution $u_{[3]}$ admits four asymptotic soliton states:
\begin{subequations}
\label{3asy1}
\begin{align}
&  u^{(1)}_{[3]} = \rho e^{2i\rho^2 t+i \phi }\left[1-\frac{8\sqrt{5}W_1(x,t)+4\left(3-\sqrt{5}\sigma\right)(\rho s_1^*-i\sigma)}{4\sqrt{5}W_1(x,t)+\left(3-\sqrt{5}\sigma\right)(2\rho s_1^*-i\sigma)}\right],\label{3asy1a}\\[2.2mm]
& u^{(2)}_{[3]} =\rho e^{2i\rho^2 t+i \phi }\left[1-\frac{8\sqrt{5}W_2(x,t)-4\left(3+\sqrt{5}\sigma\right)(\rho s_1^*-i\sigma)}{4\sqrt{5}W_2(x,t)-\left(3+\sqrt{5}\sigma\right)(2\rho s_1^*-i\sigma)}\right],\label{3asy2a}\\[2.2mm]
& u^{(3)}_{[3]} =\rho e^{2i\rho^2 t+i \phi }\left[1-\frac{8\sqrt{5}W_3(x,t)+4\left(3-\sqrt{5}\sigma\right)\rho s_1}{4\sqrt{5}W_3(x,t)+\left(3-\sqrt{5}\sigma\right)(2\rho s_1+i\sigma)}\right],\label{3asy3a} \\[2.2mm]
& u^{(4)}_{[3]} =\rho e^{2i\rho^2 t+i \phi }\left[1-\frac{8\sqrt{5}W_4(x,t)-4\left(3+\sqrt{5}\sigma\right)\rho s_1}{4\sqrt{5}W_4(x,t)-\left(3+\sqrt{5}\sigma\right)(2\rho s_1+i\sigma)}\right],\label{3asy4a}
\end{align}
\end{subequations}
with
\begin{subequations}
\begin{align}
W_1=t\eta^{-2}-\left(\frac{\sqrt{5}}{10}-\frac{\sigma}{6}\right)\rho\eta, \quad
W_2 = t\eta^{-2}+\left(\frac{\sqrt{5}}{10}+\frac{\sigma}{6}\right)\rho\eta, \label{3W12} \\ W_3=t\xi^{-2}+\left(\frac{\sqrt{5}}{10}-\frac{\sigma}{6}\right)\rho\xi, \quad
W_4=t\xi^{-2}-\left(\frac{\sqrt{5}}{10}+\frac{\sigma}{6}\right)\rho\xi. \label{3W34}
\end{align}
\end{subequations}
Moreover, the center trajectories of asymptotic solitons $u^{(i)}_{[3]}$ ($1\leq i\leq 4$) are given by the algebraic curves
\begin{subequations}
\begin{align}
& \mathcal{C}^{(1)}_{[3]}:\, t=\left(\frac{\sqrt{5}}{10}-\frac{\sigma}{6}\right)\rho\left(\eta^3 -3 s_{1R}\eta^2\right), \quad \,\,\,\, \mathcal{C}^{(2)}_{[3]}:\,
t= -\left(\frac{\sqrt{5}}{10}+\frac{\sigma}{6}\right)\rho\left(\eta^3 - 3 s_{1R}\eta^2\right),  \label{Traj3a} \\
& \mathcal{C}^{(3)}_{[3]}:\, t=-\left(\frac{\sqrt{5}}{10}-\frac{\sigma}{6}\right)\rho\left(\xi^3 + 3 s_{1R} \xi^2\right), \quad
\mathcal{C}^{(4)}_{[3]}:\, t=\left(\frac{\sqrt{5}}{10}+\frac{\sigma}{6}\right)\rho\left(\xi^3 + 3 s_{1R} \xi^2\right),  \label{Traj3b}
\end{align}
\end{subequations}
and $u_{[3]}$ approaches the asymptotic solitons $u^{(i)}_{[3]}$ respectively along the curves $\mathcal{C}^{(i)}_{[3]}$ ($1\leq i\leq 4$)  at the rate of $\mathcal{O}(t^{-1/3})$.
\end{theorem}

\beginproof
For case (i) in~\eref{AsyRel3}, dividing the numerator and denominator of Eq.~\eref{3medisolutiona} simultaneously by $\xi^6\eta^5$ and replacing $t$ via $t\sim   \Big(\!\pm\!\frac{\sqrt{5}}{10}-\frac{\sigma}{6}\Big)\rho \eta^{3} + W_{1,2}(x,t)\eta^2 $, we have
\begin{align}
u_{[3]} = \rho e^{2i\rho^2 t+i \phi }\left[1-\frac{8\sqrt{5}\,W_{1,2}(x,t) + 4(\pm 3-\sqrt{5}\sigma)(\rho s_1^*-i\sigma)+ \mathcal{O}(\eta^{-1})}{4\sqrt{5}\,W_{1,2}(x,t)
+ (\pm 3-\sqrt{5}\sigma)(2\rho s_1^*-i\sigma)+\mathcal{O}(\eta^{-1})}\right],  \label{3asy1ab1}
\end{align}
where $W_{1,2}$ are given in Eq.~\eref{3W12}, and the subscripts ``1'' and ``2'' correspond to the signs ``$+$'' and ``$-$'', respectively. Then, by calculating the Taylor series expansion of Eq.~\eref{3asy1ab1} in terms of $\eta^{-1}$ and replacing  $\mathcal{O}(\eta^{-1})$ by $\mathcal{O}(t^{-\frac{1}{3}})$ through~\eref{AsyRel3}, we obtain that
\begin{align}
u_{[3]} \sim \rho e^{2i\rho^2 t+i \phi }\left[1-\frac{8\sqrt{5}\,W_{1,2}(x,t) + 4(\pm 3-\sqrt{5}\sigma)(\rho s_1^*-i\sigma)}{4\sqrt{5}\,W_{1,2}(x,t) + (\pm 3-\sqrt{5}\sigma)(2\rho s_1^*-i\sigma)} +\mathcal{O}(t^{-\frac{1}{3}}) \right], \label{3asy1ab}
\end{align}
which shows that $u_{[3]}$ tends to the asymptotic solitons~\eref{3asy1a} and~\eref{3asy2a} (corresponding to the ``$+$'' and ``$-$'' signs of case (i) in~\eref{AsyRel3}, respectively) at the rate of $\mathcal{O}(t^{-1/3})$.

Likewise, for case (ii) in~\eref{AsyRel3}, we divide the numerator and denominator of Eq.~\eref{3medisolutionb} simultaneously by $\xi^5\eta^6$ and calculate the Taylor series expansion of $u_{[3]}$ in terms of $\xi^{-1}$, yielding
\begin{align}
u_{[3]} \sim \rho e^{2i\rho^2 t+i \phi}\left[1-\frac{8\sqrt{5}W_{3,4}(x,t) + 4\left(\pm 3-\sqrt{5}\sigma\right)\rho s_1}{4\sqrt{5}W_{3,4}(x,t)
+\left(\pm3-\sqrt{5}\sigma\right)(2\rho s_1+i\sigma)}+\mathcal{O}(t^{-1/3}) \right],  \label{3asy1cd}
\end{align}
where $\mathcal{O}(\xi^{-1})$ has been replaced by $\mathcal{O}(t^{-\frac{1}{3}})$ through~\eref{AsyRel3}, $W_{3,4}$ are given in Eq.~\eref{3W34}, and the subscripts ``3'' and ``4'' correspond to the signs ``$+$'' and ``$-$'', respectively. Accordingly, Eq.~\eref{3asy1cd} implies that $u_{[3]}$ tends to the asymptotic solitons~\eref{3asy3a} and~\eref{3asy4a} (which are associated with the ``$+$'' and ``$-$'' signs of case (ii) in~\eref{AsyRel3}, respectively) at the rate of $\mathcal{O}(t^{-1/3})$.

Moreover, the extreme value analysis shows that $\big|u^{(i)}_{[3]}\big|^2$ ($1\leq i \leq 4$) has a unique maximum/mininum for $W_i\in(-\infty, \infty)$ ($1\leq i \leq 4$), and the locations of those extrema are obtained as follows:
\begin{subequations}
\label{ExtrPoints3}
\begin{align}
&W_1(x,t)=\frac{-3+\sqrt{5}\sigma}{2\sqrt{5}}\rho s_{1R}, \quad  W_2(x,t)=\frac{3+\sqrt{5}\sigma}{2\sqrt{5}}\rho s_{1R}, \\ & W_3(x,t)=\frac{-3+\sqrt{5}\sigma}{2\sqrt{5}}\rho s_{1R}, \quad W_4(x,t)=\frac{3+\sqrt{5}\sigma}{2\sqrt{5}}\rho s_{1R},
\end{align}
\end{subequations}
which exactly define the soliton center trajectories as given in Eqs.~\eref{Traj3a} and~\eref{Traj3b}.
\hfill \endproof

\subsection{Fourth-order rational solution}
\label{Sec3.4}

The fourth-order rational solution $u_{[4]}$ can be obtained by substituting Eqs.~\eref{A1} into solution~\eref{NPT} with $N=4$. Here, we  also omit the explicit expression of $u_{[4]}$ for saving space.
Unlike the second- and third-order rational solutions, the property (i) in Lemma~\ref{lemma2} does not apply to $u_{[4]}$ but property (ii) still holds true. That is to say, it is possible that  $|\xi|, |\eta| \ra \infty$ or  $|\xi|, |\eta| =\mathcal{O}(1)$ along the soliton center trajectories $\mathcal{C}$ as $|t| \ra \infty$. In view of the relation $\xi-\eta=4\sigma\rho t$, there are only four possible cases for the asymptotic behavior of $\xi$ and $\eta$:
(a) $|\xi|=\mathcal{O}(|t|),\, |\eta|=\mathcal{O}(1)$; (b) $|\xi|=\mathcal{O}(|t|),\, |t|=\mathcal{O}(|\eta|^\alpha)$ ($\alpha >1$); (c) $|\eta|=\mathcal{O}(|t|),\, |\xi|=\mathcal{O}(1)$; (d) $|\eta|=\mathcal{O}(|t|),\, |t|=\mathcal{O}(|\xi|^\beta)$ ($\beta>1$).

For cases (a) and (b), because of $|\xi|\gg |\eta|$, we express the fourth-order rational solution in the form
\begin{align}
& u_{[4]}=\rho e^{2i\rho^2 t+i \phi }\left\{1 - \frac{10\xi^{10}\left(945t^3 + 63\rho^2t\eta^6 - 2\sigma\rho^3\eta^9\right)+\cdots}
{\begin{aligned}
& \big[ i \rho \xi^{10}\eta\left(9450\sigma t^3+90\sigma\rho^2 t\eta^6-2\rho^3\eta^9 \right) - 5i\sigma\xi^{10}\!\left(2\rho s_1^*-i\sigma\right)  \\
& \,\times \left(945 t^3 + 63 \rho^2 t\eta^6 - 2\sigma\rho^3\eta^9 \right)+\cdots \big] \end{aligned}
}\right\}, \label{4solution1}
\end{align}
where the dots stands for all the lower-power terms of $\xi$, $\eta$ and $t$.
If $|\xi|=\mathcal{O}(|t|)$ and $|\eta|=\mathcal{O}(1)$,  $\xi^{10} t^3$ is the leading term in the numerator,
whereas $\xi^{10}\eta t^3 $ and $\xi^{10} t^3$ are both the leading ones in the denominator. As a result,
the limit of Eq.~\eref{4solution1} yields a non-plane-wave asymptotic state which corresponds to an asymptotic soliton.
If $|\xi|=\mathcal{O}(|t|)$ and $|t|=\mathcal{O}(|\eta|^{\alpha})$ ($\alpha>1$), we should remove the most dominant contribution from $i\rho\xi^{10}\eta(9450\sigma t^3+90\sigma\rho^2 t\eta^6-2\rho^3\eta^{9})$ (which is greater than all the terms in the numerator) to avoid that the asymptotic limit of $u_{[4]}$ is a plane wave. For doing so, we assume that $t$ and $\eta$ obey the following asymptotic relation:
\begin{align}
t\sim V(x,t) \eta^{\alpha} + W(x,t)\eta^{\gamma} \quad (\alpha>\gamma>0),  \label{AsyRe41}
\end{align}
where $V(x,t), W(x,t)=\mathcal{O}(1)$, $\alpha$ and $\gamma$ are two real constants to be determined. Looking at the relation~\eref{AsyRe41}, we obtain the asymptotic limits of $u_{[4]}$ when $|t|\to \infty$ as follows:
\begin{align}
& u_{[4]} \sim \left\{
\begin{array}{c}
\rho e^{2i\rho^2 t+i \phi },\,\,\,\,\,\,\,\,\,\,\,\,\,\,\,\,\,\,\,\,\,\,\,\,
t\sim V(x,t)\eta^\alpha + W(x,t)\eta^{\gamma}\,\,\,(V\neq \sigma \rho V_0\,\, \text{or}\,\,\alpha\neq 3), \\[3mm]
\rho e^{2i\rho^2 t+i \phi},
\,\,\,\,\,\,\,\,\,\,\,\,\,\,\,\,\,\,\,\,\,\,\,\,\,\,\,\,\,\,\,\,\,\,\,\,\,\,\,\,\,\,\,\,\,\,\,\,\,\,\,\,\,
t\sim \sigma \rho V_0\eta^3+W(x,t)\eta^{\gamma}\,\,\,(2<\gamma<3),\\[3mm]
\rho e^{2i\rho^2 t+i \phi }\left(1 +\frac{2}{2i\sigma\rho s_1^* + 1}\right),\,\,\,\,\,\,\,\,\,\,\,
t\sim \sigma \rho V_0\eta^3+W(x,t)\eta^{\gamma}\,\,\,(0<\gamma<2),\\[3mm]
\rho e^{2i\rho^2 t+i \phi }\left[1-\frac{6i\sigma V_0}{2\sigma W(x,t) + 3V_0(2\rho s_1^*-i\sigma)}\right],\,\,\,\,\,\,\,\,\,\,\,\, t\sim \sigma \rho V_0\eta^3+W(x,t)\eta^2,
\end{array}  \right.  \label{4balancec1}
\end{align}
with
\begin{equation}
V_0 = \frac{2^{\frac{1}{3}}\left[(3\sqrt{21}+7)^{\frac{1}{3}} - (3\sqrt{21}-7)^{\frac{1}{3}}\right]}{3\times 70^{\frac{2}{3}}}.  \label{parameter1}
\end{equation}
Here, one can note that only the last case in~\eref{4balancec1} characterizes a non-plane-wave asymptotic state, which implies that a second asymptotic soliton of $u_{[4]}$ is available with the asymptotic relation $t\sim \sigma \rho V_0\eta^3 + W(x,t)\eta^{2}$ ($W=\mathcal{O}(1)$).

Similarly for cases (c) and (d),  we write the fourth-order rational solution in another form
\begin{align}
u_{[4]}=\rho e^{2i\rho^2 t+i \phi}
\left\{ 1+\frac{10\eta^{10}\left(945t^3+63\rho^2 t \xi^6+2\sigma\rho^3\xi^9\right)+\cdots}{
\begin{aligned}
& \big[i\rho\xi\eta^{10}\left(9450\sigma t^3+90\sigma\rho^2 t \xi^6+2\rho^3\xi^9\right)
+5i\sigma\eta^{10}\!\left(2\rho s_1+i\sigma\right) \\
& \,\times \left(945t^3+63\rho^2 t \xi^6+2\sigma\rho^3\xi^9\right)+\cdots \big]\end{aligned}   }
\right\},\label{4solution2}
\end{align}
where the dots stands for all the lower-power terms of $\xi$, $\eta$ and $t$.
Likewise, if $|\eta|=\mathcal{O}(|t|)$ and $|\xi|=\mathcal{O}(1)$,
the limit of Eq.~\eref{4solution2} gives a non-plane-wave asymptotic state which corresponds to a third asymptotic soliton of $u_{[4]}$.  If $|\eta|=\mathcal{O}(|t|)$ and $|t|=\mathcal{O}(|\xi|^{\beta})$ ($\beta>1$), we also need to remove the most dominant contribution from $i\rho\xi\eta^{10}\!\left(9450\sigma t^3+90\sigma\rho^2 t \xi^6+2\rho^3\xi^9\right)$, so that the asymptotic limit of $u_{[4]}$ is not a plane wave. Then, we assume that $t$ and $\xi$ satisfy the following asymptotic relation:
\begin{align}
t\sim V(x,t)\xi^\beta + W(x,t)\xi^{\delta} \quad (\beta > \delta > 0),   \label{AsyRe42}
\end{align}
where $V(x,t), W(x,t)=\mathcal{O}(1)$, $\beta$ and $\delta$ are two real constants to be determined.  Keeping in mind the relation~\eref{AsyRe42}, we obtain the asymptotic limits of $u_{[4]}$ when $|t|\to \infty$ as follows:
\begin{align}
& u_{[4]} \sim \left\{
\begin{array}{c}
\rho e^{2i\rho^2 t+i \phi },\,\,\,\,\,\,\,\,\,\,\,\,\,\,\,\,\,\,\,\,\,\,
t\sim V(x,t)\xi^\beta + W(x,t)\eta^{\delta}\,\,\,(V\neq -\sigma \rho V_0\,\, \text{or}\,\,\beta\neq 3), \\[3mm]
\rho e^{2i\rho^2 t+i \phi },\,\,\,\,\,\,\,\,\,\,\,\,\,\,\,\,\,\,\,\,\,\,\,\,\,\,\,\,\,\,\,\,\,\,\,\,\,\,\,\,\,\,\,\,\,\,\,\,\,\,\,\,
t\sim -\sigma \rho V_0 \xi^3+W(x,t)\xi^{\delta}\,\,\,(2<\delta<3),\\[3mm]
\rho e^{2i\rho^2 t+i \phi }\left(1 +\frac{2}{2i\sigma\rho s_1-1}\right),\,\,\,\,\,\,\,\,\,\,\,
t\sim -\sigma \rho V_0 \xi^3+W(x,t)\xi^{\delta}\,\,\,(0<\delta<2),\\[3mm]
\rho e^{2i\rho^2 t+i \phi }\left[1-\frac{6i\sigma V_0}{2\sigma W(x,t)
+3 V_0 (2\rho s_1+i \sigma)}\right],\,\,\,\,\,\,\,\,\,\,\, t\sim -\sigma \rho V_0 \xi^3+W(x,t)\xi^2,
\end{array}  \right.  \label{4balancec2}
\end{align}
with $V_0$ given by~\eref{parameter1}. Again, only the last case in~\eref{4balancec2} describes a non-plane-wave asymptotic state, which means that a fourth asymptotic soliton of $u_{[4]}$ is available with the asymptotic relation $t\sim -\sigma \rho V_0 \xi^3 + W(x,t)\xi^2$ ($W=\mathcal{O}(1)$).

Therefore, we obtain the asymptotic relations that $\xi$, $\eta$ and $t$ must satisfy when the asymptotic solitons of $u_{[4]}$ appear, as stated in the following proposition:

\begin{proposition}
As $|t|\ra\infty$, the asymptotic solitons of the fourth-order rational solution $u_{[4]}$ are formed only when $\xi$, $\eta$ and $t$ satisfy any of the asymptotic relations:
\begin{equation}
\label{4asy}
\begin{aligned}
&(i)\, |\xi|=\mathcal{O}(|t|),\, |\eta|=\mathcal{O}(1),\quad \,\,\,\,
(ii)\, |\xi|=\mathcal{O}(|t|), \, t\eta^{-2} - \sigma \rho V_0 \eta = \mathcal{O}(1),\\
&(iii)\, |\eta|=\mathcal{O}(|t|),\, |\xi|=\mathcal{O}(1),\quad
(iv)\, |\eta|=\mathcal{O}(|t|), \, t\xi^{-2} + \sigma\rho V_0 \xi = \mathcal{O}(1),
\end{aligned}
\end{equation}
where $V_0$ is defined in~\eref{parameter1}.
\end{proposition}

On this basis, we can further derive the explicit expressions for all asymptotic solitons of $ u_{[4]}$ as well as their center trajectories.

\begin{theorem}
Asymptotically as $t \to \pm\infty $, the fourth-order rational solution $u_{[4]}$ admits four asymptotic soliton states:
\begin{subequations}
\label{4asy1}
\begin{align}
& u_{[4]}^{(1)}= \rho e^{2i\rho^2 t+i \phi }\left[1+\frac{2 i \sigma}{2\rho W_1(x,t) - 2\rho s_1^*+i \sigma}\right],\label{4asysoliton1}\\[3mm]
& u_{[4]}^{(2)}=\rho e^{2i\rho^2 t+i \phi }\left[1-\frac{6i\sigma V_0}{2\sigma W_2(x,t)
+3 V_0(2\rho s_1^*-i\sigma)}\right],\label{4asysoliton2}\\[3mm]
& u_{[4]}^{(3)}= \rho e^{2i\rho^2 t+i \phi }\left[1-\frac{2 i \sigma}{2\rho W_3(x,t) + 2\rho s_1 + i \sigma}\right],\label{4asysoliton3}\\[3mm]
& u_{[4]}^{(4)}=\rho e^{2i\rho^2 t+i \phi }\left[1-\frac{6i\sigma V_0}{2\sigma W_4(x,t)
+3 V_0(2\rho s_1+i \sigma)}\right],\label{4asysoliton4}
\end{align}
\end{subequations}
with
\begin{align}
W_1(x,t)=\eta,\,\,\, W_2(x,t)=t\eta^{-2}- \sigma \rho V_0 \eta, \,\,\, W_3(x,t)=\xi, \,\,\,  W_4(x,t)=t\xi^{-2}+ \sigma \rho V_0 \xi, \label{4W12}
\end{align}
where $V_0$ is defined in~\eref{parameter1}. Moreover, the center trajectories of asymptotic solitons $u^{(i)}_{[4]}$ ($1\leq i\leq 4$) are given by two straight lines
\begin{align}
& \mathcal{C}^{(1)}_{[4]}:\, x-2\sigma\rho t=s_{1R},\quad  \mathcal{C}^{(3)}_{[4]}:\,  x+2\sigma\rho t=-s_{1R}, \label{Traj41}
\end{align}
and two algebraic curves
\begin{align}
&  \mathcal{C}^{(2)}_{[4]}:\,  t= \sigma \rho V_0 \left(\eta^3-3 s_{1R}\eta^2\right), \quad
\mathcal{C}^{(4)}_{[4]}:\,  t= - \sigma \rho V_0 \left(\xi^3 + 3 s_{1R}\xi^2\right);  \label{Traj42}
\end{align}
and $u_{[4]}$ approaches the asymptotic solitons $u^{(1,3)}_{[4]}$ respectively along the lines $\mathcal{C}^{(1,3)}_{[4]}$ at the rate of $\mathcal{O}(t^{-1})$, and approaches the asymptotic solitons $u^{(2,4)}_{[4]}$ respectively along the curves $\mathcal{C}^{(2,4)}_{[4]}$ at the rate of $\mathcal{O}(t^{-\frac{1}{3}})$.
 \end{theorem}
\beginproof
Note that the terms $\xi^{10}t^2$, $\eta^{10}t^2$, $\xi^{9}\eta t^3$, $\xi \eta^{9} t^3$, $\xi^{9}t^3$ and $\eta^{9}t^3$ are present in both the numerator and denominator of $u_{[4]}$. Thus,  we divide the numerator and denominator of $u_{[4]}$ by $\xi^{10} t^3$ and $\eta^{10} t^3$ respectively for cases (i) and (iii) in~\eref{4asy}, giving that
\begin{align}
u_{[4]}&=\rho e^{2i\rho^2 t+i \phi }\left[1+\frac{2 i \sigma + \mathcal{O}(t^{-1})}{2\rho(\eta-s_1^*)+i \sigma +\mathcal{O}(t^{-1}) }\right],  \label{4asy1a} \\
u_{[4]}& =\rho e^{2i\rho^2 t+i \phi }\left[1-\frac{2 i \sigma +\mathcal{O}(t^{-1})}{2\rho(\xi+s_1)+i \sigma +\mathcal{O}(t^{-1})}\right],   \label{4asy3a}
\end{align}
where $\mathcal{O}(\xi^{-1})$ and $\mathcal{O}(\eta^{-1})$ have been replaced by $\mathcal{O}(t^{-1})$ respectively for~\eref{4asy1a} and~\eref{4asy3a}. Then, by calculating the Taylor series expansion of Eqs.~\eref{4asy1a} and~\eref{4asy3a} in terms of $t^{-1}$, we have
\begin{align}
u_{[4]}&\sim \rho e^{2i\rho^2 t+i \phi }\left[1+\frac{2 i \sigma}{2\rho(\eta-s_1^*)+i \sigma}+\mathcal{O}(t^{-1})\right], \label{4asy1b} \\
u_{[4]}& \sim \rho e^{2i\rho^2 t+i \phi }\left[1-\frac{2 i \sigma}{2\rho(\xi+s_1)+i \sigma}+\mathcal{O}(t^{-1})\right],
 \label{4asy3b}
\end{align}
which show that $u_{[4]}$ tends to the asymptotic solitons~\eref{4asysoliton1} and~\eref{4asysoliton3}  at the rate of $\mathcal{O}(t^{-1})$, corresponding to cases (i) and (iii) in~\eref{4asy}, respectively.

For cases (ii) and (iv) in~\eref{4asy}, we can respectively write $u_{[4]}$ in the following equivalent forms:
\begin{align}
u_{[4]}&=\rho e^{2i\rho^2 t+i \phi }\left[1 + \frac{2 i \rho^3 \left(945  V_0^3 +63  V_0 -2\right)+\mathcal{O}(\eta^{-1})}{18\rho^3\left( 315 V_0^2+ 1\right) W_2(x,t)
- \sigma \rho^3 \left(2\rho s_1^*-i\sigma\right)\left(945 V_0^3 +63 V_0 -2\right)+\mathcal{O}(\eta^{-1})}\right],  \label{4asy2a} \\
u_{[4]}&=\rho e^{2i\rho^2 t+i \phi }\left[1+\frac{2 i  \rho^3 \left(945  V_0^3 +
63 V_0-2\right)+\mathcal{O}(\xi^{-1})}{18\rho^3\left(315  V_0^2+ 1\right)W_4(x,t) - \sigma\rho^3\left(2\rho s_1+i\sigma\right)\left(945 V_0^3 + 63 V_0 - 2\right)+\mathcal{O}(\xi^{-1})}\right],
\label{4asy4a}
\end{align}
where the first equation has been obtained by dividing the numerator and denominator of Eq.~\eref{4solution1} simultaneously by $\xi^{10}\eta^9$ and replacing $t$ via $t\sim \sigma \rho V_0\eta^3+W_2(x,t)\eta^2$ ($W_2=\mathcal{O}(1)$), whereas the second one has been obtained by dividing the numerator and denominator of Eq.~\eref{4solution2} simultaneously by $\xi^9\eta^{10}$ and replacing $t$ via $t\sim -\sigma \rho V_0 \xi^3+W_4(x,t)\xi^2$ ($W_4=\mathcal{O}(1)$). Then, we perform the Taylor series expansion in terms of $\eta^{-1}$ and $\xi^{-1}$ respectively for Eqs.~\eref{4asy2a} and~\eref{4asy4a}, and obtain that
\begin{align}
u_{[4]} \sim \rho e^{2i\rho^2 t+i \phi }\left[1-\frac{6i\sigma V_0}{2\sigma W_2(x,t)+3V_0(2\rho s_1^*-i\sigma)}+\mathcal{O}(t^{-\frac{1}{3}})\right], \label{4asy2b} \\
u_{[4]} \sim \rho e^{2i\rho^2 t+i \phi }\left[1-\frac{6i\sigma V_0}{2\sigma W_4(x,t)+3V_0(2\rho s_1+i \sigma)}+\mathcal{O}(t^{-\frac{1}{3}})\right],  \label{4asy4b}
\end{align}
where $\mathcal{O}(\eta^{-1})$ and $\mathcal{O}(\xi^{-1})$ have been replaced by $\mathcal{O}(t^{-\frac{1}{3}})$, and $W_{2,4}$ are defined in Eq.~\eref{4W12}.  Therefore, $u_{[4]}$ tends to the asymptotic solitons~\eref{4asysoliton2} and~\eref{4asysoliton4}  at the rate of $\mathcal{O}(t^{-\frac{1}{3}})$, which are respectively associated with cases (ii) and (iv) in~\eref{4asy}.

Moreover, via the  extreme value analysis, we find that $\big|u_{[4]}^{(i)}\big|^2$ has a unique maximum/mininum for $W_i\in(-\infty, \infty)$ ($1\leq i \leq 4$), and the locations of those extrema are obtained as follows:
\begin{equation}
\begin{aligned}
W_1=s_{1R}, \,\,\, W_2=-3 \sigma\rho V_0 s_{1R}, \,\,\, W_3=-s_{1R}, \,\,\,
W_4=-3\sigma\rho V_0 s_{1R},
\end{aligned}\label{ExtrPoints4}
\end{equation}
which exactly define the soliton center trajectories as given in Eqs.~\eref{Traj41} and~\eref{Traj42}.
\hfill \endproof

\begin{figure}[H]
\centering
\subfigure[]{\label{Fig1a}
\includegraphics[width=6.5in]{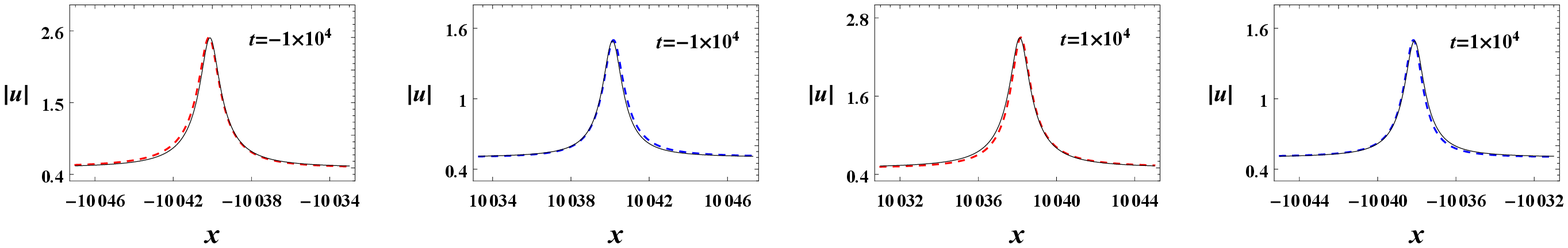}}\\[3mm]
\subfigure[]{\label{Fig1b}
\includegraphics[width=6.5in]{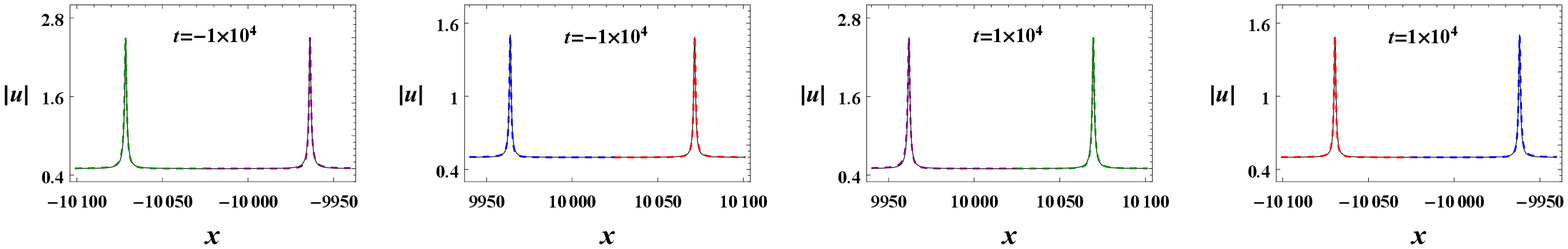}}\\[3mm]
\subfigure[]{\label{Fig1c}
\includegraphics[width=6.5in]{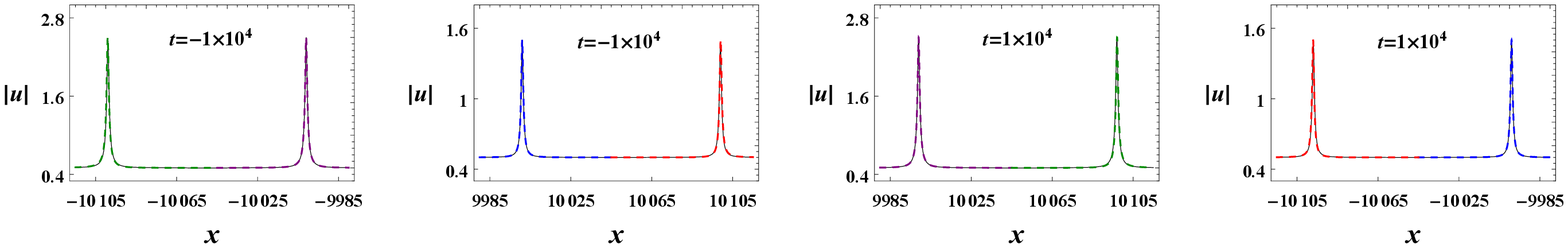}}
\caption{(a) Comparison of the asymptotic solitons $u^{(1)}_{[2]}$ (blue dashed) and $u^{(2)}_{[2]}$ (red dashed) with the second-order solution (black solid), where $\rho=0.5$, $\phi=0$, $\sigma=-1$, $s_1=1+1.5i$, $s_2=1+i $. (b) Comparison of the asymptotic solitons $u^{(1)}_{[3]}$ (blue dashed), $u^{(2)}_{[3]}$ (red dashed), $u^{(3)}_{[3]}$ (purple dashed) and $u^{(4)}_{[3]}$ (green dashed) with the third-order solution (black solid), where $\rho=0.5$, $\phi=0$, $\sigma=-1$, $s_1=1+1.5i$, $s_2=1+i$, $s_3=1+i $. (c) Comparison of the asymptotic solitons $u^{(1)}_{[4]}$ (blue dashed), $u^{(2)}_{[4]}$ (red dashed), $u^{(3)}_{[4]}$ (purple dashed) and $u^{(4)}_{[4]}$ (green dashed) with the fourth-order solution (black solid), where $\rho=0.5$, $\phi=0$, $\sigma=-1$, $s_1=1+1.5i$, $s_2=1+i$, $s_3=1+i $, $s_4=2i $.  \label{Fig1} }
\end{figure}

\subsection{Comparison of asymptotic solitons with the exact solutions}

We notice that there are always  some asymptotic solitons localized in the algebraic curves for solution~\eref{NPT} with $N\ge 2$, and they are approached by the exact solutions slower than those lying in the straight lines. Thus, it is necessary to test the validity of our asymptotic analysis for the higher-order rational solutions. In doing so, we compare the exact  rational solutions with $2 \leq N\leq 4$ and their asymptotic solitons as given in Eqs.~\eref{2asy1},~\eref{3asy1}, and~\eref{4asy1} at large values of $|t|$. As shown in Fig.~\ref{Fig1}, all the asymptotic solitons have a good agreement with the exact solutions in the far-field region of the $xt$ plane, which indicates that our asymptotic analysis gives the accurate expressions of asymptotic solitons.

\section{Dynamical properties of soliton interactions}

\label{Sec4}

In this section, based on the asymptotic expressions obtained in Section~\ref{Sec3}, we will discuss the
dynamical properties of soliton interactions described by solution~\eref{NPT}.
\begin{figure}[H]
\centering
\subfigure[]{\label{f2a}
\includegraphics[width=1.9in]{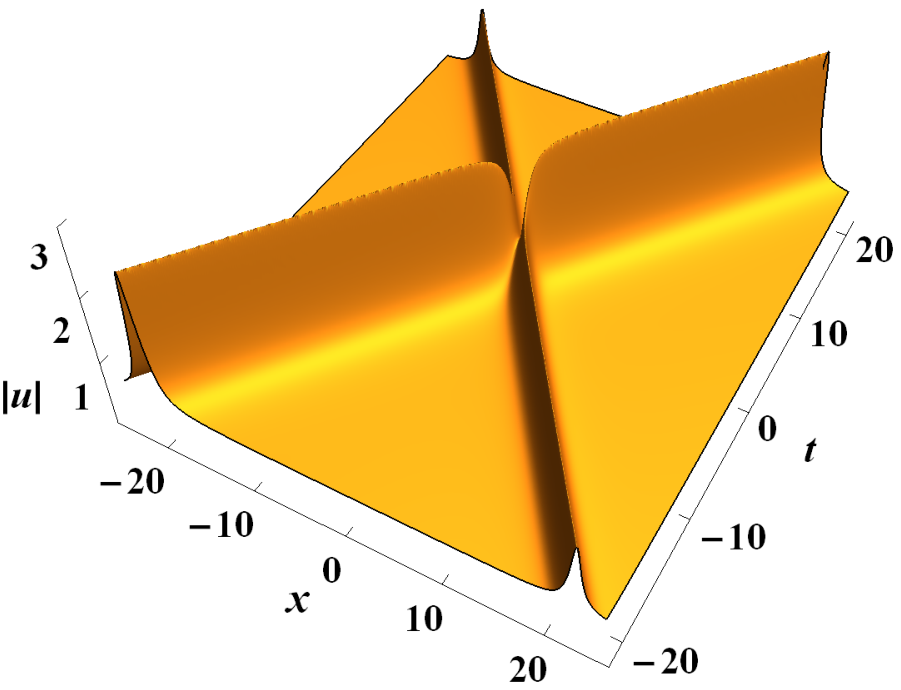}}\hfill
\subfigure[]{ \label{f2b}
\includegraphics[width=1.9in]{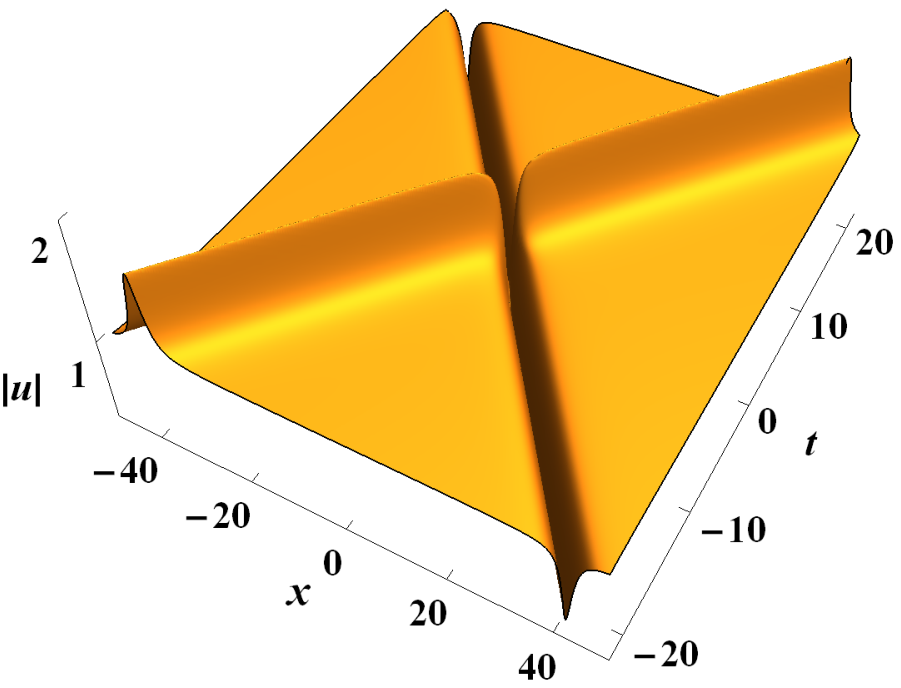}}\hfill
\subfigure[]{ \label{f2c}
\includegraphics[width=1.9in]{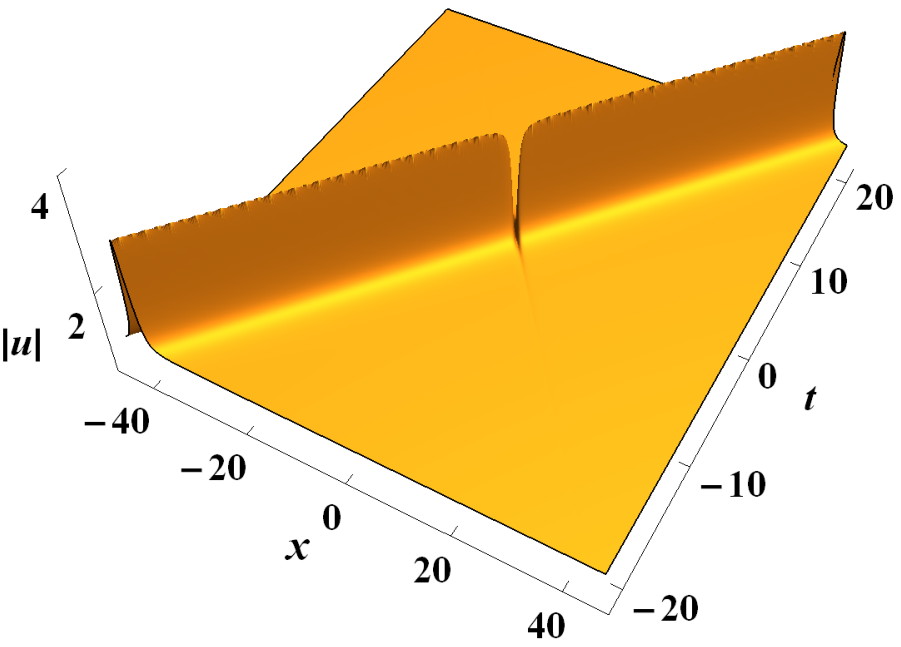}}\\
\caption{\small Interactions between two straight asymptotic solitons via solution~\eref{1solution}:
(a) RAD-RAD soliton interaction with $\rho=0.5$, $\phi=0$, $\sigma=-1$, $s_1=1+1.5 i $; (b)  RD-RAD soliton interaction with $\rho=1$, $\phi=0$, $\sigma=-1$, $s_1=-1+2 i$; (c) V-RAD soliton interaction with $\rho=1$, $\phi=0$, $\sigma=-1$,  $s_1=1+ i $.   \label{Fig2} }
\end{figure}

\begin{figure}[H]
\centering
\subfigure[]{\label{f3a}
\includegraphics[width=1.9in]{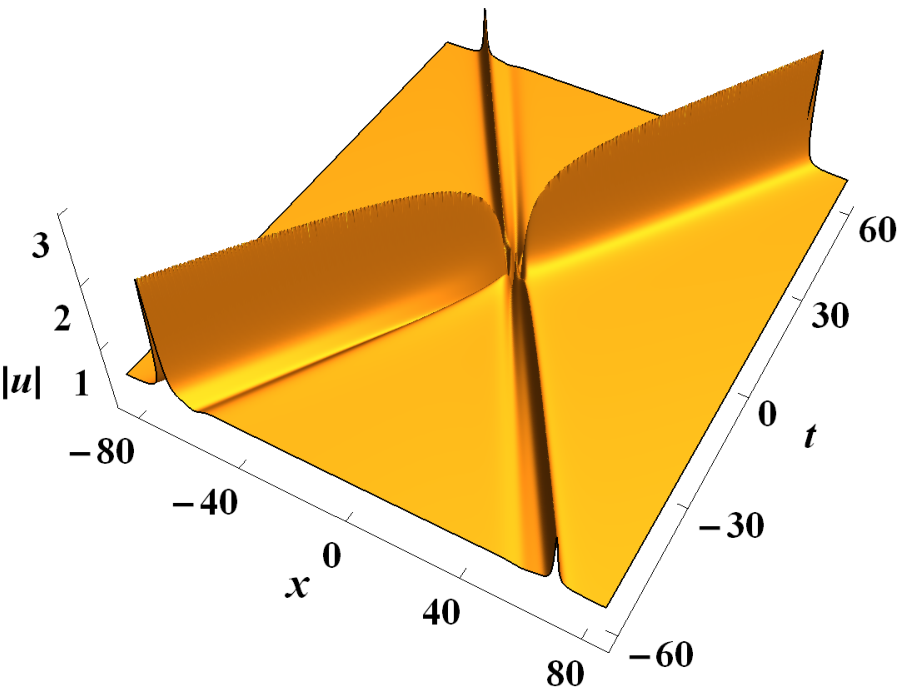}}\hfill
\subfigure[]{ \label{f3b}
\includegraphics[width=1.9in]{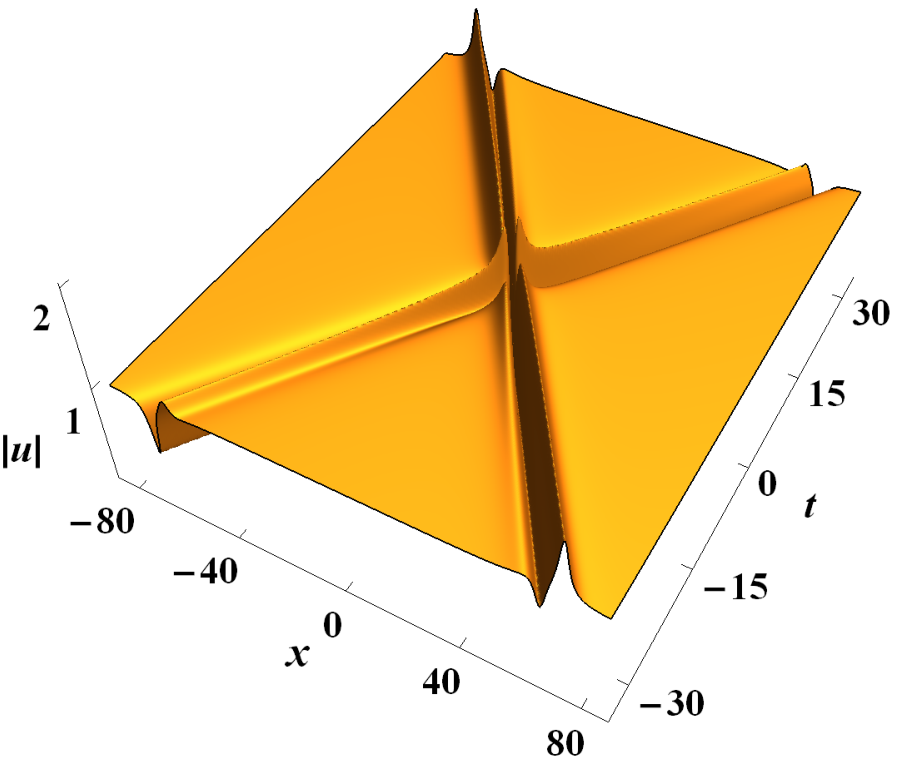}}\hfill
\subfigure[]{ \label{f3c}
\includegraphics[width=1.9in]{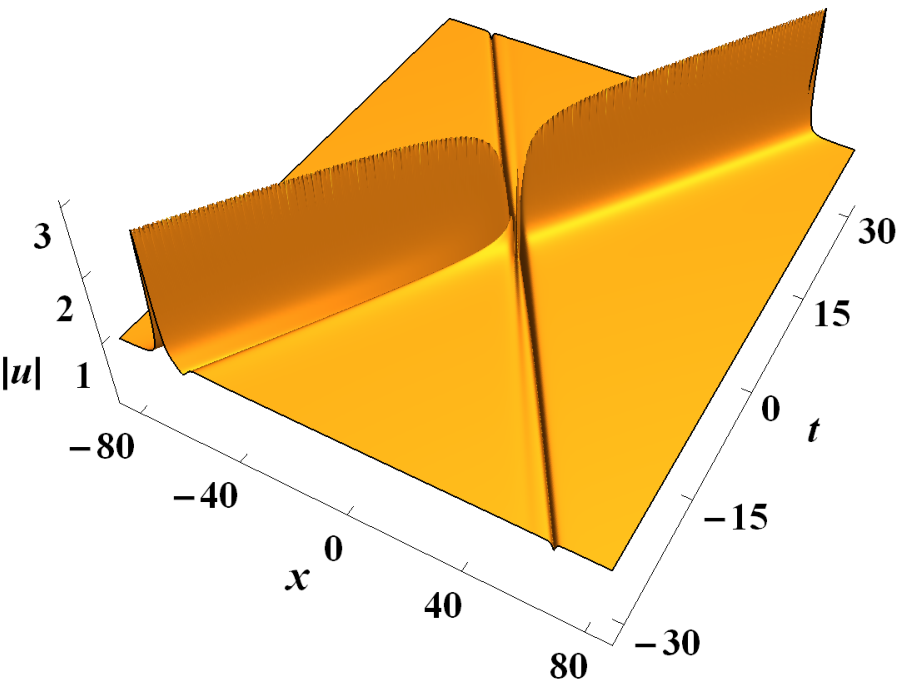}}\\
\caption{\small Interactions between two curved asymptotic solitons via solution~\eref{2solution}:  (a) RAD-RAD soliton interaction with $\rho=0.5$, $\phi=0$, $\sigma=-1$, $s_1=1+1.5 i $, $s_2=1+ i $; (b)  RAD-RD soliton interaction with $\rho=1$, $\phi=0$, $\sigma=-1$, $s_1=- i$, $s_2=1-12 i $; (c) V-RAD soliton interaction with $\rho=0.5$, $\phi=0$, $\sigma=-1$,  $s_1=1+2 i $, $s_2=0.2 i $. \label{Fig3} }
\end{figure}

\begin{figure}[H]
 \centering
\subfigure[]{\label{f4a}
\includegraphics[width=1.9in]{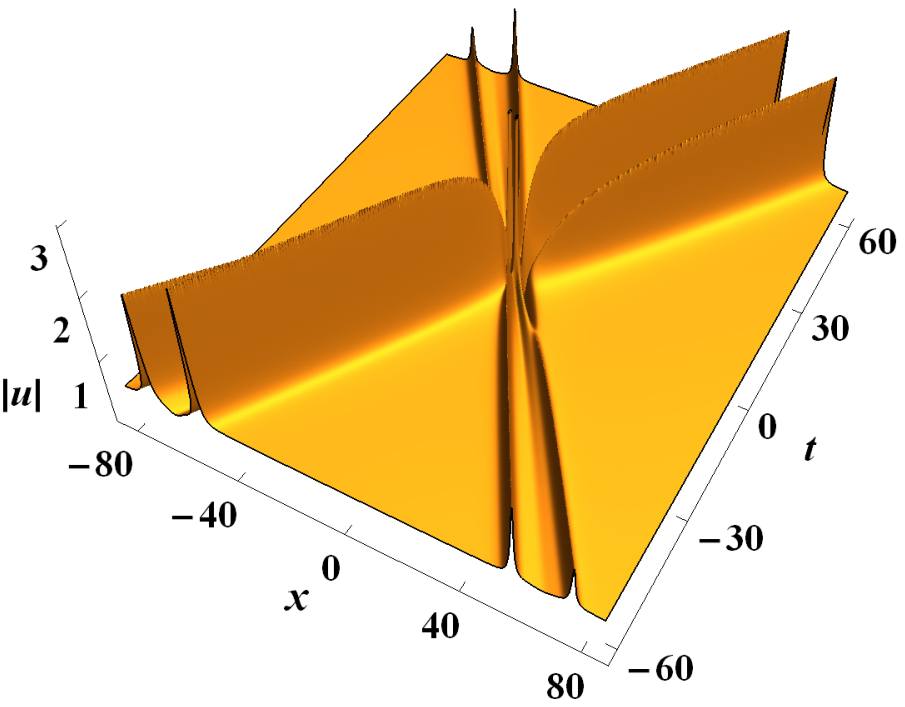}}\hfill
\subfigure[]{ \label{f4b}
\includegraphics[width=1.9in]{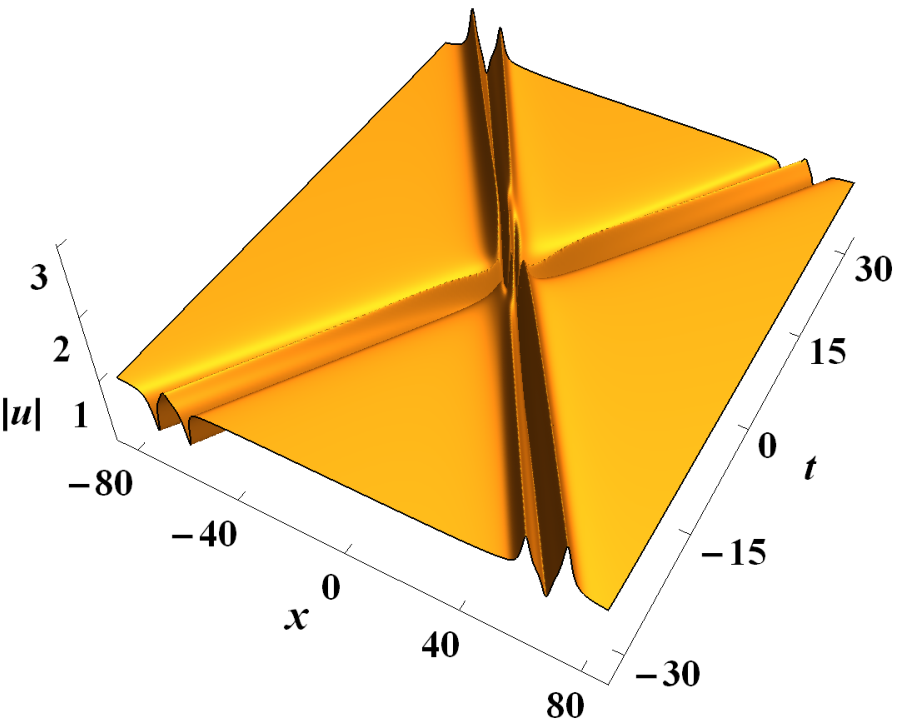}}\hfill
\subfigure[]{ \label{f4c}
\includegraphics[width=1.9in]{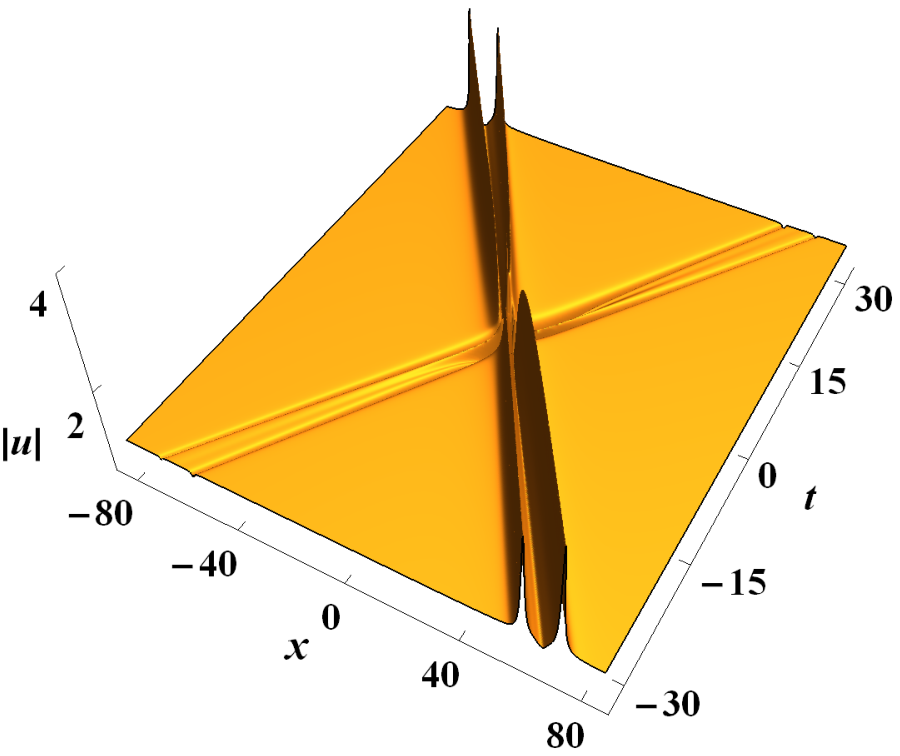}}\\
\caption{\small Interactions among four curved asymptotic solitons via solution~\eref{NPT} with $N=3$:  (a) RAD-RAD-RAD-RAD soliton interaction with $\rho=0.5$, $\phi=0$, $\sigma=-1$, $s_1=1+1.5 i $, $s_2=1+ i $, $s_3=1+5 i $; (b) RAD-RAD-RD-RD soliton interaction with $\rho=1$, $\phi=0$, $\sigma=-1$, $s_1=- i$, $s_2=1 $, $s_3=1-50 i $; (c) RAD-RAD-V-V soliton interaction with $\rho=1$, $\phi=0$, $\sigma=-1$,  $s_1=-1 $, $s_2=1 $, $s_3=1-20 i $. \label{Fig4} }
\end{figure}

\begin{figure}[H]
 \centering
\subfigure[]{\label{f5a}
\includegraphics[width=1.9in]{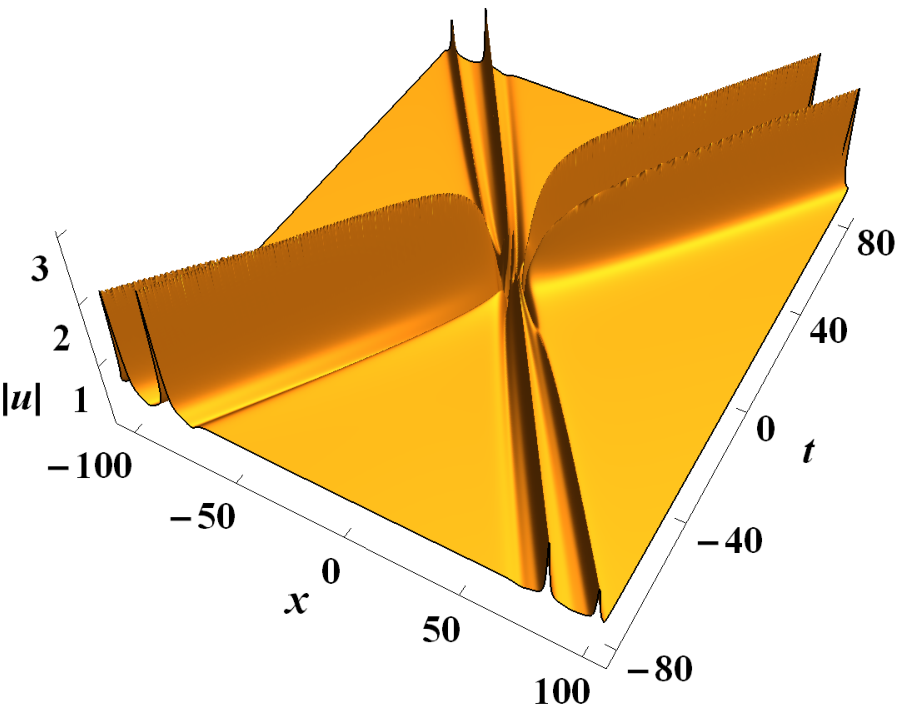}}\hfill
\subfigure[]{ \label{f5b}
\includegraphics[width=1.9in]{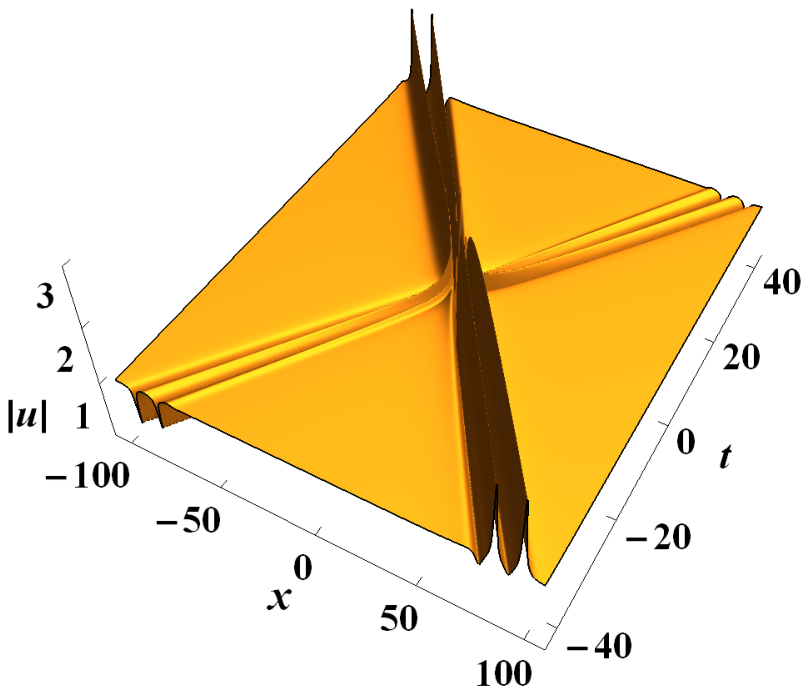}}\hfill
\subfigure[]{ \label{f5c}
\includegraphics[width=1.9in]{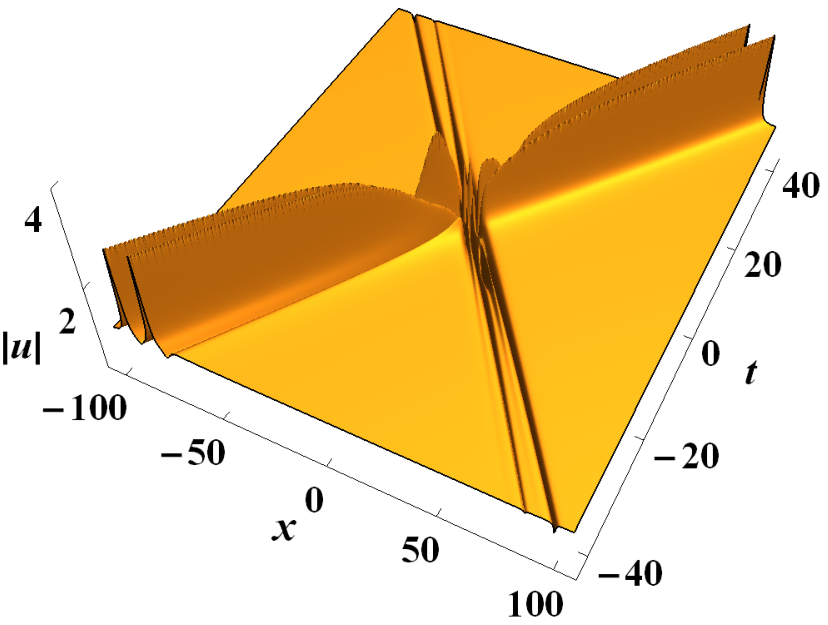}}\\
\caption{\small Interactions among two straight asymptotic solitons and two curved asymptotic solitons via solution~\eref{NPT} with $N=4$:  (a) RAD-RAD-RAD-RAD soliton interaction with $\rho=0.5$, $\phi=0$, $\sigma=-1$, $s_1=1+1.5 i $, $s_2=1+ i $, $s_3=1 $, $s_4=1+4 i $; (b)  RAD-RAD-RD-RD soliton interaction with $\rho=1$, $\phi=0$, $\sigma=-1$, $s_1=-1-0.2 i$, $s_2=1- i $, $s_3=-1-2 i $, $s_4=1-10 i $; (c) V-V-RAD-RAD soliton interaction with $\rho=1$, $\phi=0$, $\sigma=-1$,  $s_1=1+ i $, $s_2=0.5 i $, $s_3=-50+10 i $, $s_4=50+200 i $. \label{Fig5} }
\end{figure}

First, from the asymptotic expressions $u^{(i)}_{[N]}$ ($i=1,2$ for $N=1,2$ and $i=1,2,3,4$ for $N=3,4$), one can obtain that $\displaystyle\lim_{W_i\ra \pm\infty}\big|u^{(i)}_{[N]}\big|^2 = \rho^2 $ and their unique extrema take the values as follows:
\begin{equation}
\label{ExtrValues}
\begin{aligned}
& \big|u^{(1)}_{[1]}\big|^2_{\rm{ext}} =  \big|u^{(1)}_{[2]}\big|^2_{\rm{ext}} = \big|u^{(1,2)}_{[3]}\big|^2_{\rm{ext}} = \big|u^{(1,2)}_{[4]}\big|^2_{\rm{ext}} =\rho^2\left[1+\frac{8(1+\sigma\rho s_{1I})}{\left(\sigma+2\rho s_{1I}\right)^2}\right], \\
& \big|u^{(2)}_{[1]}\big|^2_{\rm{ext}} =  \big|u^{(2)}_{[2]}\big|^2_{\rm{ext}} = \big|u^{(3,4)}_{[3]}\big|^2_{\rm{ext}} = \big|u^{(3,4)}_{[4]}\big|^2_{\rm{ext}}   =\rho^2\left[1- \frac{8\sigma\rho s_{1I}}{\left(\sigma+2\rho s_{1I}\right)^2}\right]. \\
\end{aligned}
\end{equation}
It can be found that each extremum value in~\eref{ExtrValues} could be greater than, less than or equal to $\rho^2$, depending on the sign of $1+\sigma\rho s_{1I}$ or $\sigma  s_{1I}$. That is to say, all the asymptotic expressions can display both the rational dark (RD) and rational anti-dark (RAD) soliton profiles, and particularly they will disappear into the plane-wave background as $|t|\to\infty$ if $1+\sigma\rho s_{1I}=0 $ or $s_{1I}=0$. Hence, we can make a classification of the rational solutions according to the types of asymptotic solitons. It turns out that solution~\eref{NPT} with $1\leq N\leq 4 $ exhibit always five different types of soliton interactions which are respectively associated with the parametric conditions: (i) $1+\sigma\rho s_{1I}>0,\, \sigma s_{1I}<0$, (ii)  $1+\sigma\rho s_{1I}<0,\, \sigma  s_{1I}<0$, (iii) $1+\sigma\rho s_{1I}>0,\, \sigma s_{1I}>0$, (iv)  $1+\sigma\rho s_{1I}>0,\, s_{1I}=0$, (v)  $1+\sigma\rho s_{1I}=0,\, \sigma s_{1I}<0$. It should be pointed that for the third- and fourth-order rational solutions, the asymptotic solitons $u^{(1)}_{[N]}$ and $u^{(2)}_{[N]}$ ($N=3,4$) possess the same profiles, so do $u^{(3)}_{[N]}$ and $u^{(4)}_{[N]}$. To illustrate, we present some examples of the soliton interactions in Figs.~\ref{Fig3}--\ref{Fig5}, where ``V'' means the vanishment of some asymptotic soliton(s) as $|t|\to \infty$.

\begin{figure}[H]
\begin{minipage}[t]{0.48\linewidth} 
\centering
{\includegraphics[width=2.2in]{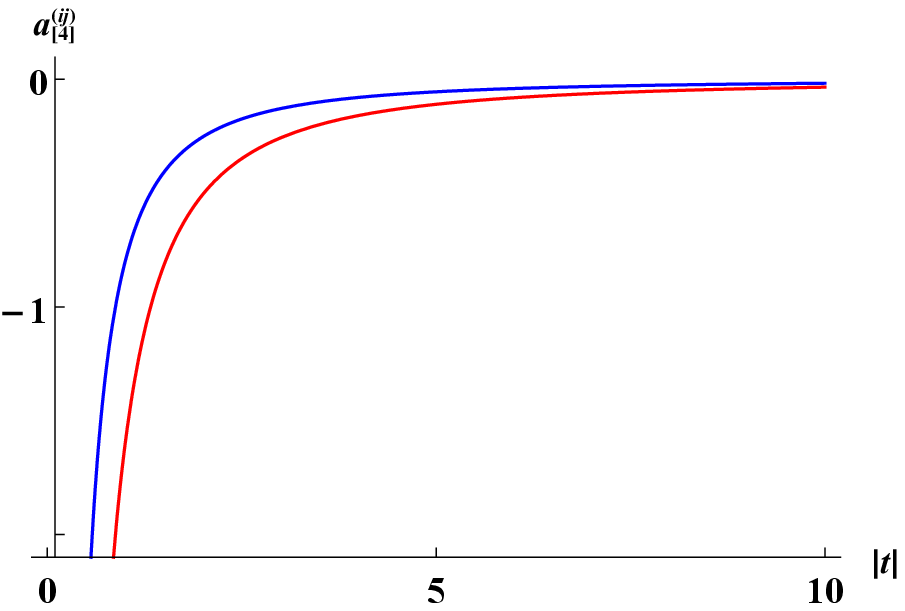}} \\
\caption{\small Two-soliton interaction forces versus $|t|$ for solution $u_{[4]}$ with $\sigma=-1$, $\rho=1$ and $s_{1R}=0$, where the red line represents the accelerations $a_{[4]}^{(12)}$, $a_{[4]}^{(14)}$, $a_{[4]}^{(23)}$ and $a_{[4]}^{(34)}$, while the blue line represents the acceleration $a^{(24)}_{[4]}$. \label{Fig6} }
\end{minipage}\hspace{0.5cm}
\begin{minipage}[t]{0.48\linewidth}
\centering
\includegraphics[width=2.3in]{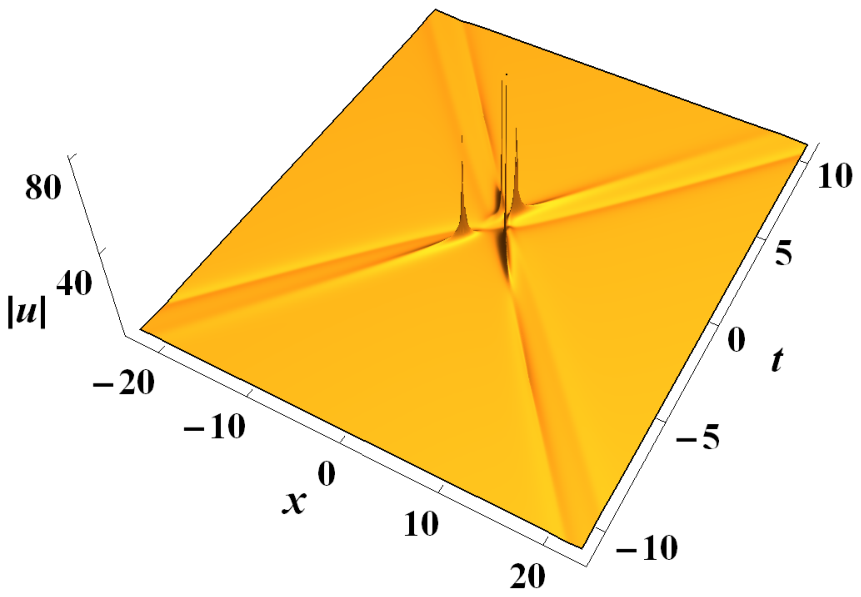}
\caption{\small Singular behavior in the second-order rational solution~\eref{2solution} with $\rho=1$, $\phi=0$, $\sigma=-1$,  $s_1=1-0.8 i $, $s_2=-0.01i $, where the singularity occurs at the four points $(x,t) \approx (\pm 2.32905,-0.9185)$ or $(\pm 0.809477,1.10721)$. \label{Fig7} }
\end{minipage}
\end{figure}

Next, in order to reveal some unusual behavior of soliton interactions in solution~\eref{NPT}, we make a quantitative analysis of asymptotic solitons $u^{(i)}_{[N]}$ ($i=1,2$ for $N=1,2$ and $i=1,2,3,4$ for $N=3,4$) in the following aspects:

\begin{enumerate}

\item[(i)]

Since every $u^{(i)}_{[N]}$ represents a pair of asymptotic solitons as $t\to \infty$ and $t\to -\infty$ simultaneously, all interacting solitons can retain their shapes and amplitudes upon mutual interactions. By calculating the absolute differences between $\big|u^{(i)}_{[N]}\big|^2_{\rm{ext}}$ and $\rho^2$, we obtain the amplitudes for $\big|u^{(i)}_{[N]}\big|^2$ as follows:
\begin{equation}
\begin{aligned}
& A^{(1)}_{[1]} = A^{(1)}_{[2]} = A^{(1,2)}_{[3]}  = A^{(1,2)}_{[4]}  =  \frac{8\rho^2 |1+\sigma\rho s_{1I}|}{\left(\sigma+2\rho s_{1I}\right)^2}, \\
& A^{(2)}_{[1]} = A^{(2)}_{[2]}  = A^{(3,4)}_{[3]}  = A^{(3,4)}_{[4]}  = \frac{8\rho^3 |s_{1I}|}{\left(\sigma+2\rho s_{1I}\right)^2},
\end{aligned}
\end{equation}
which implies that  the asymptotic solitons of solution~\eref{NPT} with any given $N$ can be divided into two halves with each having the same amplitudes.

\item[(ii)]

Based on the expressions of $\mathcal{C}^{(i)}_{[N]}$, the velocities of asymptotic solitons $u^{(i)}_{[N]}$ can be given by
\begin{equation}
\begin{aligned}
& v^{(1)}_{[1]} = v^{(1)}_{[4]}= 2\sigma\rho, \qquad\qquad\qquad\qquad\qquad\,\,  v^{(2)}_{[1]} = v^{(3)}_{[4]} = -2\sigma\rho, \\
& v^{(1)}_{[2]}=2\sigma\rho+\frac{1}{\sigma\rho\left(\eta^2-2 s_{1R} \eta\right)},
\quad\quad\quad\quad\quad v^{(2)}_{[2]}=-2\sigma\rho-\frac{1}{\sigma\rho\left(\xi^2+2 s_{1R} \xi\right)}, \\
& v^{(1,2)}_{[3]} = 2\sigma\rho+\frac{1}{\frac{\pm 3\sqrt{5}-5\sigma}{10}\rho\left(\eta^2-2 s_{1R} \eta\right)}, \quad
 v^{(3,4)}_{[3]}= -2\sigma\rho-\frac{1}{\frac{\pm 3\sqrt{5}-5\sigma}{10}\rho\left(\xi^2 + 2 s_{1R}\xi\right)},  \\
&v_{[4]}^{(2)}=2\sigma\rho+\frac{1}{3\sigma \rho V_0\left(\eta^2-2 s_{1R}\eta\right)},\qquad \quad\,\,\,
v_{[4]}^{(4)}=-2\sigma\rho-\frac{1}{3\sigma \rho V_0\left(\xi^2 + 2 s_{1R}\xi\right)},
\end{aligned}
\end{equation}
where $V_0$ is defined in~\eref{parameter1}, the superscripts ``1'' and ``2'' (``3'' and ``4'') respectively correspond to the signs ``+'' and ``-'' for $v^{(i)}_{[3]}$. Note that $\xi$, $\eta$ and $t$ satisfy the equations of $\mathcal{C}^{(i)}_{[N]}$, thus the above velocities apart from $v^{(1,2)}_{[1]}$ and $v^{(1,3)}_{[4]}$ are $t$-dependent and they approach the constant $2\sigma\rho$ or $-2\sigma\rho$ at the same rate $\mathcal{O}(t^{-2/3})$. Moreover, suppose that $(x_1, t)$ and $(x_2, -t)$  are two points in the curves $\mathcal{C}^{(i)}_{[N]}$ for any $t\in \mathbb{R}$. If $s_{1R}=0$, we have $x_1=-x_2$, so that $\xi(x_1, t)=-\xi(x_2, -t)$ and $\eta(x_1, t)=-\eta(x_2, -t)$. In this case, the  velocities $v^{(i)}_{[N]}$ take the same values at $t$ and $-t$. But if $s_{1R} \neq 0$, $x_1 \neq -x_2$ implies that $\xi(x_1, t) \neq -\xi(x_2, -t)$ and $\eta(x_1, t) \neq -\eta(x_2, -t)$. Accordingly, the asymptotic solitons localized in the algebraic curves have different velocities at $t$ and $-t$ when $s_{1R} \neq 0$. However, such a difference tends to $0$ as $|t|\to \infty$, which can be seen from Table~\ref{Table1}.

\item[(iii)]

The relative distance $d^{(ij)}_{[N]}$ between two asymptotic solitons $u^{(i)}_{[N]}$ and $u^{(j)}_{[N]}$ ($i\neq j$) can be obtained by calculating their absolute position difference at certain time. Then,  we use the second derivative of $d^{(ij)}_{[N]}$ with respect to $t$ (i.e., the acceleration $a^{(ij)}_{[N]}$ that two asymptotic solitons separate from each other) to measure the two-soliton interaction force.  It is found that the interaction force between two solitons with the straight center trajectories is $0$ since the relative distance is linear in $t$; otherwise, the interaction forces are of the attractive type ($a^{(ij)}_{[N]}<0$) and their strengths decay to $0$ at the rate $\mathcal{O}(t^{-5/3})$. For example, with $N=4$ and $s_{1R}=0$,  the two-soliton separation accelerations $a^{(ij)}_{[4]}$ ($1\leq i\leq j\leq 4$) are given as follows:
\begin{align}
a^{(13)}_{[4]}=0, \quad a^{(12)}_{[4]}= a^{(14)}_{[4]}= a^{(23)}_{[4]}= a^{(34)}_{[4]}= -\frac{2}{9} \sqrt[3]{\frac{1}{\rho V_0}}\,t^{-\frac{5}{3}}, \quad a^{(24)}_{[4]}=-\frac{4}{9} \sqrt[3]{\frac{1}{\rho V_0}}\,t^{-\frac{5}{3}},
\end{align}
for which we plot the variation of $a^{(ij)}_{[4]}$ with the increase of $|t|$ in Fig.~\ref{Fig6}.
\end{enumerate}

Therefore, we can see that for solution~\eref{NPT} with $N\ge 2$, the soliton interactions are completely elastic when $s_{1R}=0$ because all the solitons can retain their individual shapes, amplitudes and velocities upon mutual interactions. But if $s_{1R} \neq 0$, the soliton interactions are \emph{quasi-elastic} in the sense that there exists a slight difference for the velocities of curved asymptotic solitons between at $t$ and $-t$. Meanwhile, we should mention that the two-soliton interaction forces for the exponential and exponential-and-rational solutions are absolutely $0$ or exponentially decaying to $0$ as $|t|\to \infty$~\cite{LiXu,LiXu3}. That is to say, the soliton interactions in the rational solutions with $N\ge 2$ are stronger than those in the other two types of solutions for the defocusing NNLS equation.

\begin{table}[h]
\small
\caption{Velocities of curved asymptotic solitons with the parameters selected as $\sigma=-1$, $\rho=1$ and $s_{1R}=0.2$.}
\begin{center}
\begin{tabular}{ccccccccc}
\hline
$t$&-500&-200&-100&-50&50&100&200&500\\ \hline
$v_{[2]}^{(1)}$ &-2.0076290&-2.0140491&-2.0222937&-2.0353693&-2.0353743&-2.0222953&-2.0140496&-2.0076292\\ \hline
$v_{[2]}^{(2)}$ &2.0076290&2.0140491&2.0222937&2.0353693&2.0353743&2.0222953&2.0140496&2.0076292\\ \hline
$v_{[3]}^{(1)}$ &-1.9927618&-1.9866707&-1.9788484&-1.9664421&-1.9664477&-1.9788502&-1.9866713&-1.9927619\\ \hline
$v_{[3]}^{(2)}$ &-2.0137526&-2.0253304&-2.0402054&-2.0638113&-2.0638129&-2.0402058&-2.0253306&-2.0137526\\ \hline
$v_{[3]}^{(3)}$ &1.9927618&1.9866707&1.9788484&1.9664421&1.9664477&1.9788502&1.9866713&1.9927619\\ \hline
$v_{[3]}^{(4)}$ &2.0137526&2.0253304&2.0402054&2.0638113&2.0638129&2.0402058&2.0253306&2.0137526\\ \hline
$v_{[4]}^{(2)}$ &-2.0191118&-2.0352028&-2.0558780&-2.0886931&-2.0886939&-2.0558782&-2.0352029&-2.0191118\\ \hline
$v_{[4]}^{(4)}$ &2.0191118&2.0352028&2.0558780&2.0886931&2.0886939&2.0558782&2.0352029&2.0191118\\ \hline
\end{tabular}
\end{center} \label{Table1}
\end{table}

\section{Conclusions and discussions}

\label{Sec5}

In this paper, for the defocusing NNLS equation (i.e., Eq.~\eref{NNLS} with $\varepsilon= -1$), we have constructed the $N$th-order rational solutions based on the DT and some limit technique, and have studied the asymptotic behavior and soliton interactions in the rational solutions with $1\leq N\leq 4$. Finally, we address the conclusions and discussions of this paper as follows:

First, by using the $N$-fold DT and choosing the plane-wave solution~\eref{cws} as a seed, we have obtained the $N$th-order rational solutions expressed as the ratio of two determinants (see Eq.~\eref{NPT}), in which the elements can be determined recursively. Note that the rational solutions are just some degenerate cases of the exponential solutions when all the spectral parameters coalesce at the critical value $\lam= i\sigma \rho$ ($\sigma=\pm1$). In our derivation, we have employed some limit technique to deal with the coalescence of multiple spectral parameters, which has been widely used in constructing the rogue-wave solutions of integrable models~\cite{Guo}. Compared with the previous paper~\cite{LiXu1}, we have given a rigorous proof on the determinant representation of the $N$th-order rational solutions.

Second, we have obtained the explicit expressions of all asymptotic solitons (which are localized in the straight or curved lines) for solution~\eref{NPT} with $1\leq N \leq 4$. The key point of our asymptotic analysis is to find the balances between $t$ and $\xi$ or between $t$ and $\eta$ up to the subdominant level. Note that all the rational solutions are expressible in terms of $\xi$, $\eta$ and $t$.
Therefore, we can in principle make an asymptotic analysis of solution~\eref{NPT} for any given $N$ by following the procedure: (i) find the asymptotic behavior of $\xi$ and $\eta$ as $|t|\to\infty$ on basis of the relation $\xi-\eta=4\sigma\rho t$; (ii) determine the asymptotic relations among $\xi$, $\eta$ and $t$ when solution~\eref{NPT} admits the non-plane-wave limits; (iii) derive the explicit expressions of asymptotic solitons with the corresponding asymptotic relations; (iv) obtain the center trajectories of asymptotic solitons by the extreme value analysis. Also, we have shown that the exact solutions approach the asymptotic solitons localized in the algebraic curves with a slower rate than those in the straight lines, and all the asymptotic solitons have a good agreement with the exact solutions when $|t|\gg 1$.

Third, we have studied the dynamical properties of soliton interactions based on the obtained asymptotic expressions.
It turns out that all the rational solutions exhibit just five different types of soliton interactions,
and the interacting solitons are divided into two halves with each having the same amplitudes. With the given order $N\ge 2$, some or all asymptotic solitons of solution~\eref{NPT} are localized in the algebraic curves, so that their velocities are $t$-dependent and tend to $2\sigma\rho$ or $-2\sigma\rho$ at the rate $\mathcal{O}(t^{-2/3})$. Particularly, we have revealed the \emph{quasi-elastic} behavior of curved asymptotic solitons, that is, their velocities take different values between at $t$ and $-t$ but such a difference decays to $0$ as $|t|\to \infty$. Moreover, we have found that the separation accelerations between two curved solitons or between a curved soliton and a straight soliton change with $t$ in an algebraic manner, which means that the rational solutions exhibit stronger soliton interactions  than the exponential and exponential-and-rational solutions do.

In addition, we should emphasize that all the asymptotic expressions are globally nonsingular with  the condition
\begin{align}
s_{1I} \neq -\frac{\sigma }{2 \rho }.\label{singular condition}
\end{align}
However, it does not mean that the rational solutions themselves have no
singularity with the same condition. In fact, this is a necessary but not sufficient condition for solution~\eref{NPT} with $N\ge 2$ to be nonsingular. To illustrate,  Fig.~\ref{Fig7} illustrates that the rational solutions may exhibit the singular behavior in the near-field region although their asymptotic expressions always show the soliton profiles. Therefore, multiple solitons from the remote past may develop into a singularity when they interact, or a singularity may develop into several   stable solitons in the remote future, which is quite different from the usual soliton interactions in the local integrable models.

\section*{Acknowledgments}
This work was partially supported by the National Natural Science Foundation of China (Grant Nos. 11705284 and 11971322), by the Fundamental Research Funds of the Central Universities (Grant No. 2017MS051), and by the program of China Scholarship Council (Grant No. 201806445009). T. X. appreciates the hospitality of the Department of Mathematics \& Statistics at McMaster University during his visit in 2019.

\newpage
\begin{appendix}
\section{Expansion coefficients $f^{(j-1,m-1)} $ and $g^{(j-1,m-1)}$ ($1\leq j \leq 4$, $1\leq m \leq 5$)}
\renewcommand{\theequation}{A.\arabic{equation}}
\setcounter{equation}{0} \label{appendixA}
\begin{subequations}
\label{A1}
\begin{align}
& 
f^{(0,m-1)}=(i \sigma \rho)^{m-1} (i\rho\xi^{\prime}) J_{+},\quad
g^{(0,m-1)}=(i \sigma \rho)^{m-1} (i+\sigma\rho\xi^{\prime })J_{-}  \quad (1\leq m \leq 5), \label{A1a}
\\
& f^{(1,0)}=\frac{1}{4}f^{(0,0)}+\frac{1}{3}\left(i\rho\xi^{\prime }\right)^3J_{+}+i\rho T J_{+},\\
& g^{(1,0)}=\frac{1}{4}g^{(0,0)}-\frac{1}{3}\left(i+\sigma\rho\xi^{\prime }\right)^3J_{-}+\left(\sigma\rho T-\frac{i}{3}\right) J_{-}, \\
& f^{(2,0)}=\left(i\rho\xi^{\prime }\right)^2f^{(1,0)}-\frac{3}{10}\left(i\rho\xi^{\prime }\right)^4f^{(0,0)}+i\rho \left(-\frac{1}{32}\xi^{\prime }+\frac{1}{4}T+s_3\right)J_{+},\\
& g^{(2,0)}=-\left(i+\sigma\rho\xi^{\prime }\right)^2g^{(1,0)}-\frac{3}{10}\left(i+\sigma\rho\xi^{\prime }\right)^4g^{(0,0)} \notag \\
& \,\,\,\,\,\,\,\,\,\,\,\,\,\,\,\,\,\,\,\, +\left[\sigma\rho \left(-\frac{1}{32}\xi^{\prime }+\frac{1}{4}T+s_3\right)+\frac{3i}{160}\right]J_{-}, \\
&f^{(3,0)}=\left(i\rho\xi^{\prime }\right)^2f^{(2,0)}-\frac{5}{6}\left(i\rho\xi^{\prime }\right)^4f^{(1,0)}+\frac{31}{126}\left(i\rho\xi^{\prime }\right)^6f^{(0,0)}\notag\\
&\,\,\,\,\,\,\,\,\,\,\,\,\,\,\,\,\,\,\,\,-i\rho^3\xi^{\prime }\left(T+\frac{1}{4}\xi^{\prime }\right)^2J_{+}+i\rho\left(\frac{1}{128}\xi^{\prime }-\frac{1}{32}T+\frac{1}{4}s_3+s_4\right)J_{+},\\
&g^{(3,0)}=-\left(i+\sigma\rho\xi^{\prime }\right)^2g^{(2,0)}-\frac{5}{6}\left(i+\sigma\rho\xi^{\prime }\right)^4g^{(1,0)}-\frac{31}{126}\left(i+\sigma\rho\xi^{\prime }\right)^6g^{(0,0)}\notag\\
&\,\,\,\,\,\,\,\,\,\,\,\,\,\,\,\,\,\,\,\,-\left(i+\sigma\rho\xi^{\prime }\right)\left[\frac{i}{12}\sigma-\rho\left(T+\frac{1}{4}\xi^{\prime }\right)\right]^2J_{-} \notag  \\
& \,\,\,\,\,\,\,\,\,\,\,\,\,\,\,\,\,\,\,\, + \left[\sigma\rho\left(\frac{1}{128}\xi^{\prime }-\frac{1}{32}T+\frac{1}{4}s_3+s_4\right)-\frac{5i}{896}\right]J_{-},\\
& f^{(j-1,m-1)}=(i \sigma \rho)^{m-1} f^{(j-1,0)}+ \sum_{k=0}^{j-2} (-1)^{j+k} \frac{(m+j-3-k)!}{(j-k-1)!(m-2)!} f^{(k,m-1)}  \notag \\
& \hspace{7.5cm} (2\leq j \leq 4,\, 2\leq m \leq 5), \label{RecRele}
\\
& g^{(j-1,m-1)}=(i \sigma \rho)^{m-1} g^{(j-1,0)}+ \sum_{k=0}^{j-2} (-1)^{j+k} \frac{(m+j-3-k)!}{(j-k-1)!(m-2)!} g^{(k,m-1)}  \notag \\
&  \hspace{7.5cm} (2\leq j \leq 4,\, 2\leq m \leq 5),  \label{RecRelf}
\end{align}
\end{subequations}
with $ J_{+}=2\sqrt{2} e^{i(\rho^2 t+ \frac{1}{2} \phi)}$, $J_{-}=2\sqrt{2} e^{-i(\rho^2 t+ \frac{1}{2} \phi)}$, $\xi^{\prime }=x+2\sigma\rho t+s_1$ and $T=2\sigma\rho t+s_2$.
\end{appendix}

\end{document}